\documentclass[apj]{emulateapj}

\newcommand{\hii}{\textrm{H}~\textsc{ii}}

\newcommand{\oiv}{[\textrm{O}~\textsc{iv}]}

\newcommand{\neii}{[\textrm{Ne}~\textsc{ii}]}
\newcommand{\nev}{[\textrm{Ne}~\textsc{v}]}
\newcommand{\nevi}{[\textrm{Ne}~\textsc{vi}]}

\slugcomment{Accepted for publication in ApJ on 13 September 2010} 
\shorttitle{The Effect of AGNs on Aromatic Features}
\shortauthors{Diamond-Stanic et al.}

\begin{document}

\title{The Effect of Active Galactic Nuclei on the Mid-Infrared
Aromatic Features}

\author{Aleksandar M. Diamond-Stanic\altaffilmark{1}, George
H. Rieke\altaffilmark{1}}
\altaffiltext{1}{Steward Observatory, University of Arizona, Tucson,
AZ, 85721}

\begin{abstract}
We present Spitzer measurements of the aromatic (also known as PAH)
features for 35 Seyfert galaxies from the revised Shapley--Ames sample
and find that the relative strengths of the features differ
significantly from those observed in star-forming galaxies.
Specifically, the features at 6.2, 7.7, and 8.6~$\mu$m are suppressed
relative to the 11.3~$\mu$m feature in Seyferts.  Furthermore, we find
an anti-correlation between the L(7.7~$\mu$m)/L(11.3~$\mu$m) ratio and
the strength of the rotational H$_2$ emission, which traces shocked
gas.  This suggests that shocks suppress the short-wavelength features
by modifying the structure of the aromatic molecules or destroying the
smallest grains.  Most Seyfert nuclei fall on the relationship between
aromatic emission and \neii\ emission for star-forming galaxies,
indicating that aromatic-based estimates of the star-formation rate in
AGN host galaxies are generally reasonable.  For the outliers from
this relationship, which have small L(7.7~$\mu$m)/L(11.3~$\mu$m)
ratios and strong H$_2$ emission, the 11.3~$\mu$m feature still
provides a valid measure of the star-formation rate.
\end{abstract}


\section{Introduction}

The mid-infrared (mid-IR) aromatic emission features are a universal
product of star formation in moderate-to-high--metallicity galaxies
\citep[e.g.,][]{roc91,lu03,smi07a}.  Their molecular carriers, often
assumed to be polycyclic aromatic hydrocarbons
\citep[PAHs,][]{leg84,all85,tie08}, radiate through IR fluorescence
following vibrational excitation by a single ultraviolet (UV) photon
\citep{tie05} and provide an indirect measurement of the UV radiation
field strength, and therefore the star-formation rate (SFR), that is
largely extinction independent \citep[e.g.,][]{pee04,cal07,rie09}.
This emission is thought to originate in photo-dissociation regions
where aromatic molecules are shielded from the harsh radiation field
near hot stars \citep[e.g.,][]{pov07}.  These molecules can also be
destroyed by the harder radiation field associated with an active
galactic nucleus \citep[AGN,][]{voi92,gen98}.  Nonetheless, aromatic
features are readily detected in many AGNs above IR continua boosted
by hot dust, and they have been used to probe the SFR in AGN host
galaxies \citep[e.g.,][]{sch06, shi07, shi09, lut08}.  Understanding
what environments destroy or modify these features is important for
constraining systematic uncertainties in aromatic-based estimates of
the SFR, and is a key way to probe the nature of their molecular
carriers, an open issue in our understanding of the interstellar
medium.

\citet{dul81} first suggested that vibrational modes of aromatic
hydrocarbons could produce the observed features, which were
subsequently identified with specific C--H and C--C bending and
stretching modes \citep{all89}.  Specifically, the 6.2 and 7.7~$\mu$m
features are produced by C--C stretching modes, the 8.6~$\mu$m feature
by C--H in-plane bending modes, and the 11.3 and 12.7~$\mu$m features
by C--H out-of-plane bending modes.  While these features are commonly
attributed to PAHs, we use the simpler, more general term ``aromatic''
to avoid making assumptions about the detailed structure of the
molecules.  It is worth noting, for example, that PAH spectra from
laboratory measurements and quantum chemical calculations are unable
to match the range of astronomical spectra without artificial
enhancements of the 6.2, 7.7, and 8.6~$\mu$m feature strengths
\citep[e.g.,][]{li01}.  Regardless of this uncertainty associated with
uniquely matching observed spectra with expectations for specific
molecules, one can probe the properties of the aromatic carriers by
measuring the relative strengths of the emission features, which are
expected to vary as a function of charge state \citep[e.g.,][]{bak01},
molecular size \citep[e.g.,][]{dra01}, and molecular structure
\citep[e.g.,][]{ver02}.

Efforts to study variations in aromatic feature strengths outside the
Milky Way have focused on star-forming galaxies
\citep[e.g.,][]{smi07a,gal08,ros09,odo09}, but the AGNs included in
these studies have shown evidence for suppression of shorter
wavelength features (e.g., those at 6.2, 7.7, and 8.6~$\mu$m) relative
to those at longer wavelengths.  For example, \citet{smi07a} studied a
sample of 59 galaxies from the Spitzer Infrared Nearby Galaxy Survey
\citep[SINGS,][]{ken03}, of which 12 have Seyfert nuclei and 20 have
low-ionization nuclear emission-line regions (LINERs), and they found
a tendency for the systems with reduced 6--8~$\mu$m features to be
associated with low-luminosity AGNs.  They speculated on possible
causes for this behavior, including whether the AGN can modify the
aromatic grain size distribution or serve as the excitation source for
aromatic emission.  Similarly, \citet{odo09} studied a sample of 92
galaxies at $z\sim0.1$ from the Spitzer SDSS GALEX Spectroscopic
Survey, including eight AGN-dominated and 20 composite systems, and
found that the AGNs exhibited lower 7.7/11.3~$\mu$m ratios.  They
interpreted this behavior as being consistent with destruction of
small aromatic molecules by shocks or X-rays associated with the AGNs,
but they were unable to distinguish any differences between the
AGN-dominated and composite objects, nor any strong correlation with
AGN power.  The physical slit size at their median redshift is 6 kpc,
so there is little spatial information.

In this paper we analyze the aromatic features drawing from the sample
of 89 local Seyfert galaxies studied by \citet{dia09}.  This sample is
from the revised Shapley--Ames catalog \citep[RSA,][]{san87}, and
includes every galaxy brighter than $B_T=13$ that is known to host
Seyfert activity \citep{mai95,ho97}.  The median distance of the
sample is 22~Mpc, so the 3.7\arcsec~slit width of the Short-Low (SL)
module of the Infrared Spectrograph \citep[IRS,][]{hou04} on the
Spitzer Space Telescope \citep{wer04} provides spatial information on
scales of a few hundred parsecs, allowing us to isolate nuclear
regions distinct from the rest of the galaxy.  As a result, we are
able to probe the effect of AGNs on the aromatic features more
systematically than has previously been done.

\section{Data}

We gathered data from the Spitzer archive taken with the IRS SL module
($\lambda=5.2$--14.5~$\mu$m, $R=64$--128) from a variety of programs
(24, 86, 159, 668, 3247, 3269, 3374, 30471, 30572, 30577, 30745,
40018, and 50702), as well as dedicated data taken for this study
(program 40936, PI: G.~H. Rieke).  We use CUBISM \citep{smi07b} to
extract one-dimensional spectra from the basic calibrated data using
$3.6\arcsec\times7.2\arcsec$ apertures oriented along the slit
direction.  This aperture size was chosen to isolate the nuclear
component of the aromatic emission while still including a substantial
fraction of the diffraction-limited beam.  We use the calibration for
extended sources based on the assumption that the regions producing
aromatic emission are spatially extended, so the extracted spectra are
in units of surface brightness.  We use overlapping data in the
7.59--8.42~$\mu$m region to scale the SL2/SL3 orders to the SL1 order
when offsets are apparent; these offsets are $<10$\% in all cases.

\begin{figure}[!t]
\begin{center}
\includegraphics[angle=0,scale=.45]{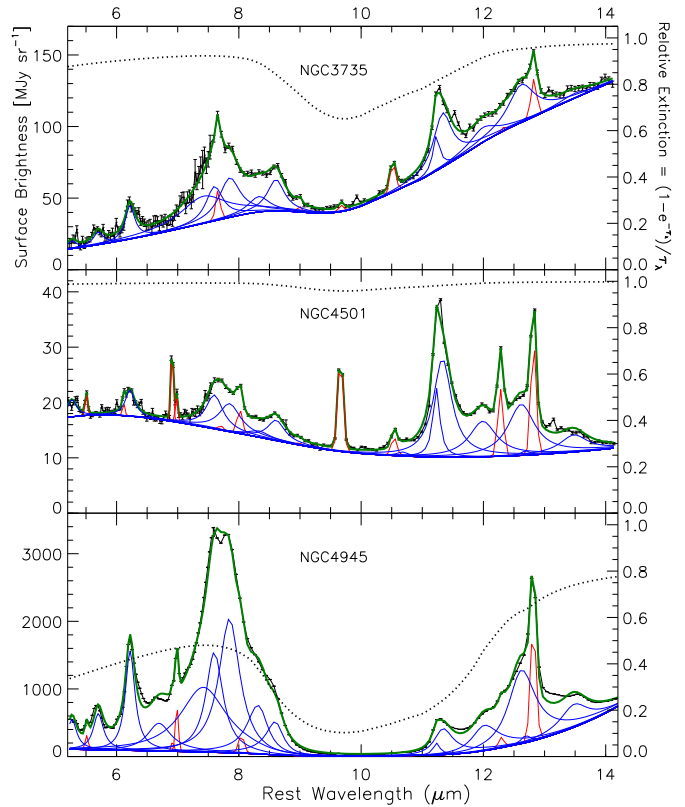}
\caption{Example PAHFIT spectral decompositions for three RSA Seyfert
  nuclei.  The observed spectra are shown in black.  The blue lines
  above the continuum correspond to the aromatic features, while red
  lines correspond to unresolved atomic and molecular emission lines.
  The total fit is shown in green, and the dotted line indicates the
  extinction profile.}
\label{fig:decomp}
\end{center}
\end{figure}

We then use a modified version of PAHFIT \citep{smi07a} to determine
the strength of the various aromatic features.  This spectral fitting
package includes aromatic features represented by Drude profiles, dust
continuum emission represented by modified blackbodies at fixed
temperatures, fine-structure lines and H$_2$ rotational lines
represented by Gaussian profiles, starlight represented by $T=5000$~K
blackbody emission, and dust extinction represented by a power-law and
silicate features.  Because Seyfert galaxies exhibit higher-ionization
emission lines, silicate dust emission, and hot-dust continuum
emission, we additionally include a \nevi~$\lambda$7.652$~\mu$m
emission line and a silicate emission component, both represented by
Gaussian profiles, and we use temperatures of 1000, 750, 500, 350,
225, 150, and 100~K for the dust continuum emission.  We show example
PAHFIT decompositions for three sources exhibiting a range in
continuum shape and silicate extinction in Figure~\ref{fig:decomp}.

\vspace{1cm}

\section{Results}\label{sec:results}

The data exhibit a range of signal-to-noise ratios (S/N), so we
visually inspected all of the nuclear spectra to identify those that
have clear detections of the relevant aromatic features and whose
spectra are adequately described by the PAHFIT model.  The spectra for
these 35 sources are shown in Figure~\ref{fig:spec}.  Since many of
the observations were executed with the mapping mode of IRS, it is
also possible to extract spectra for off-nuclear regions in some
galaxies, allowing for comparison between spectra dominated by the
active nucleus and spectra dominated by \hii\ regions within the same
galaxy.  We identified off-nuclear regions that were covered by the
IRS slit and had sufficient S/N to detect the relevant aromatic
features in 21/35 galaxies.  We show a comparison between the nuclear
and off-nuclear extractions for these galaxies in
Figure~\ref{fig:off}, and we compile relevant measurements in
Tables~\ref{tab:nuc} and \ref{tab:off}.  

\begin{figure*}[!t]
\begin{center}
\includegraphics[angle=0,scale=0.9]{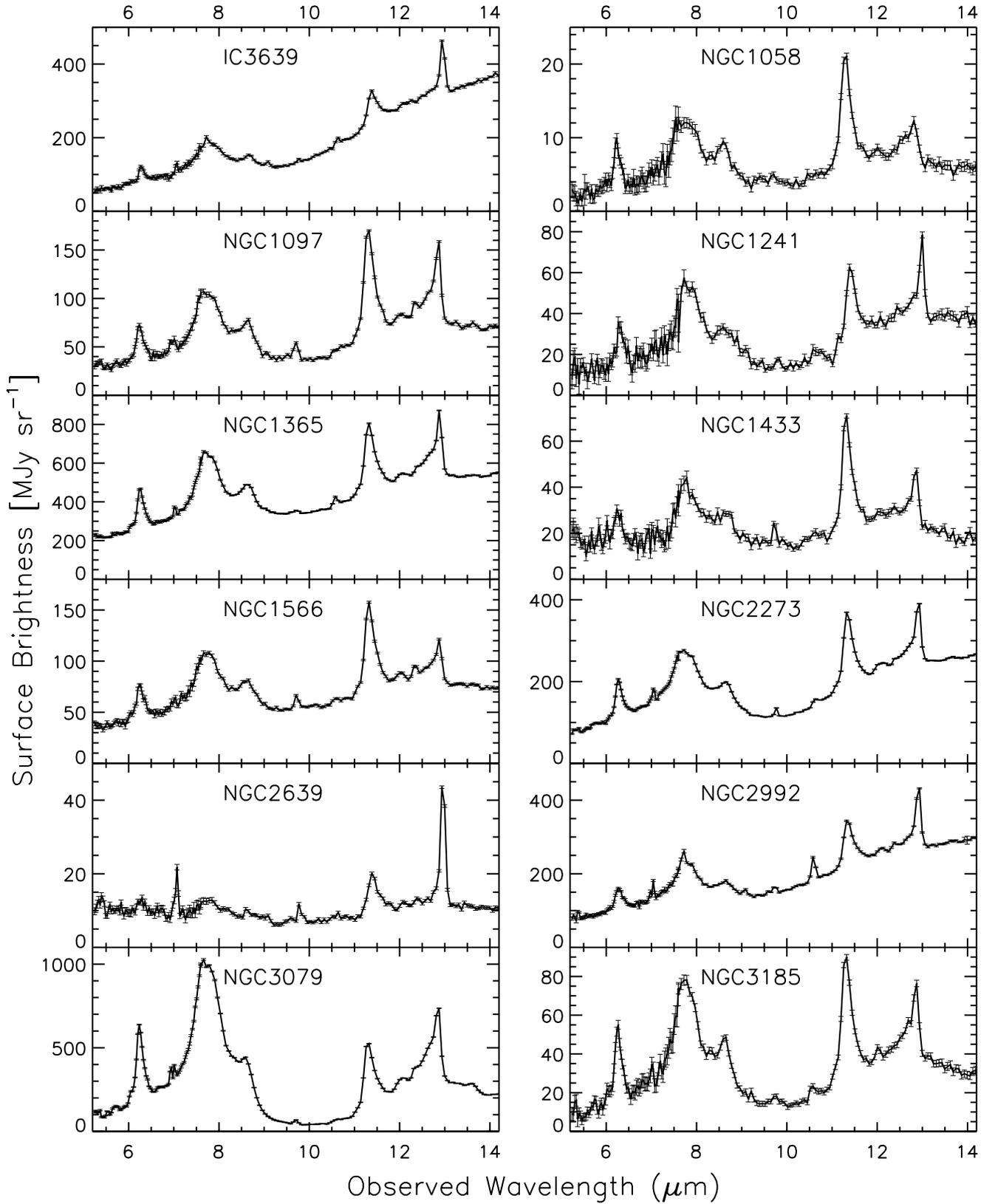}
\caption{Low-resolution 5.2--14.2~$\mu$m Spitzer/IRS nuclear spectra
  for the 35 RSA Seyfert galaxies considered in this study.}
\label{fig:spec}
\end{center}
\end{figure*}


\begin{figure*}[!h]
\figurenum{\ref{fig:spec}}
\begin{center}
\includegraphics[angle=0,scale=.9]{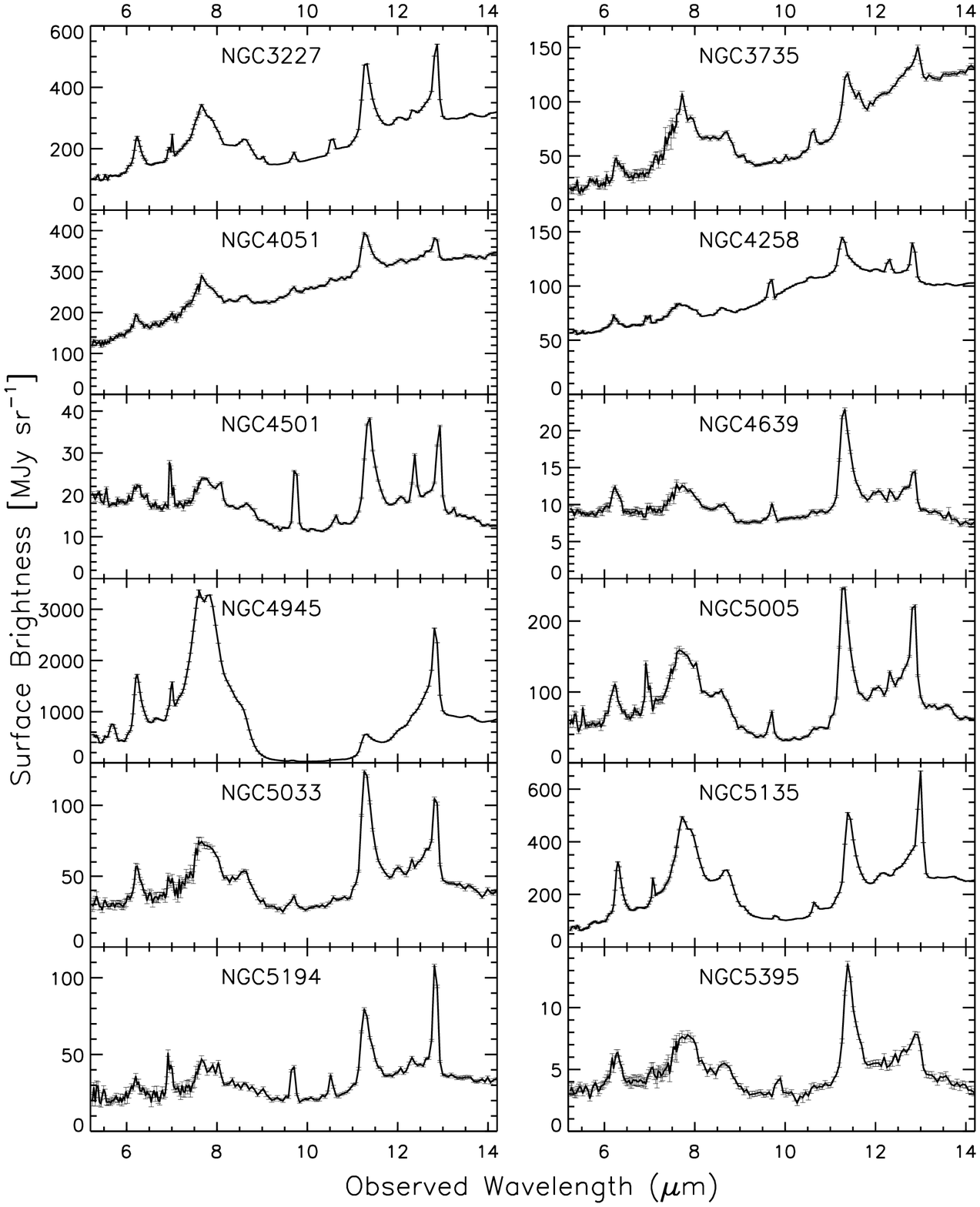}
\caption{{\it Continued}}
\label{fig:spec_p2}
\end{center}
\end{figure*}


\begin{figure*}[!h]
\figurenum{\ref{fig:spec}}
\begin{center}
\includegraphics[angle=0,scale=.9]{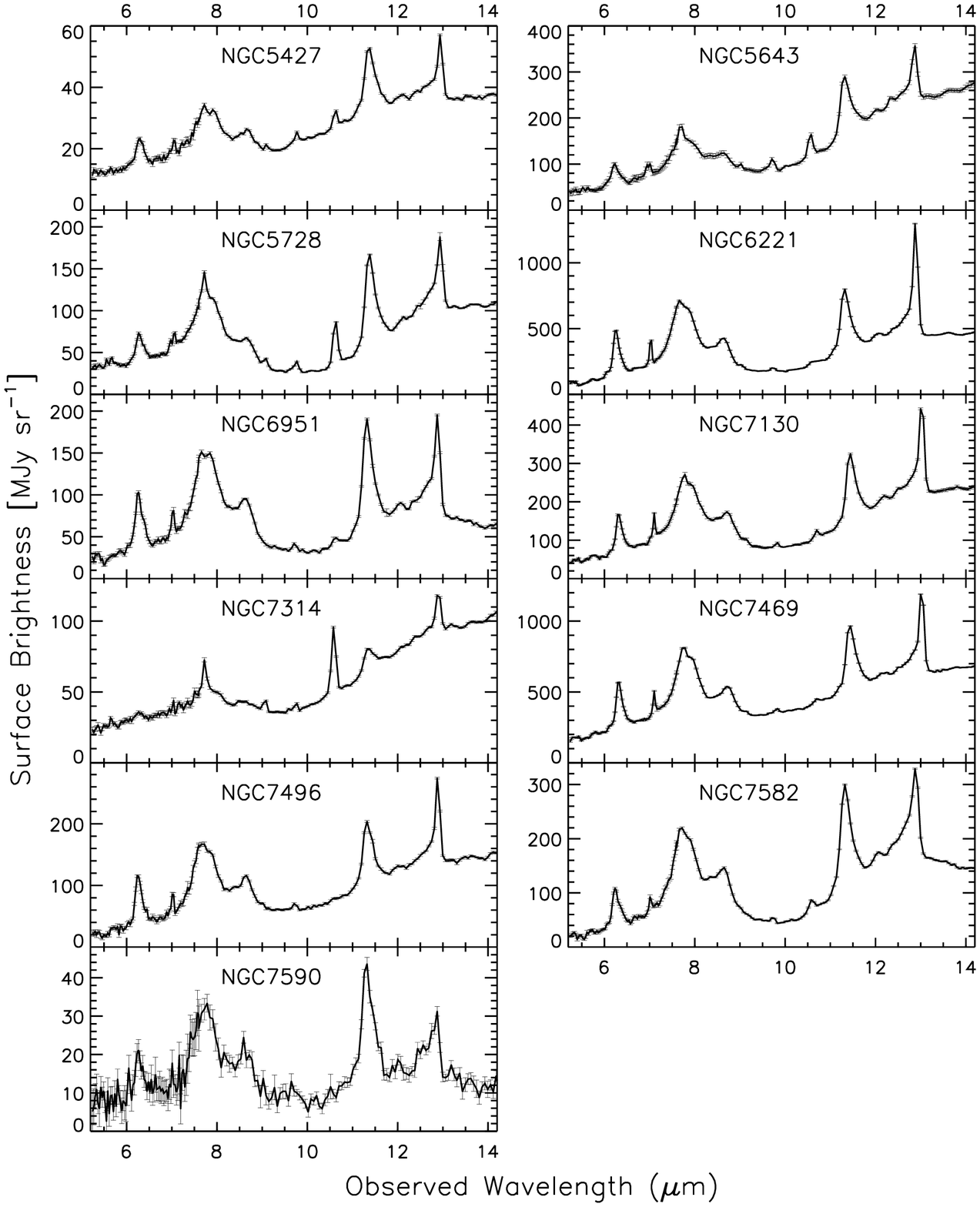}
\caption{{\it Continued}}
\label{fig:spec_p3}
\end{center}
\end{figure*}

\begin{figure*}[!t]
\begin{center}
\includegraphics[angle=0,scale=.9]{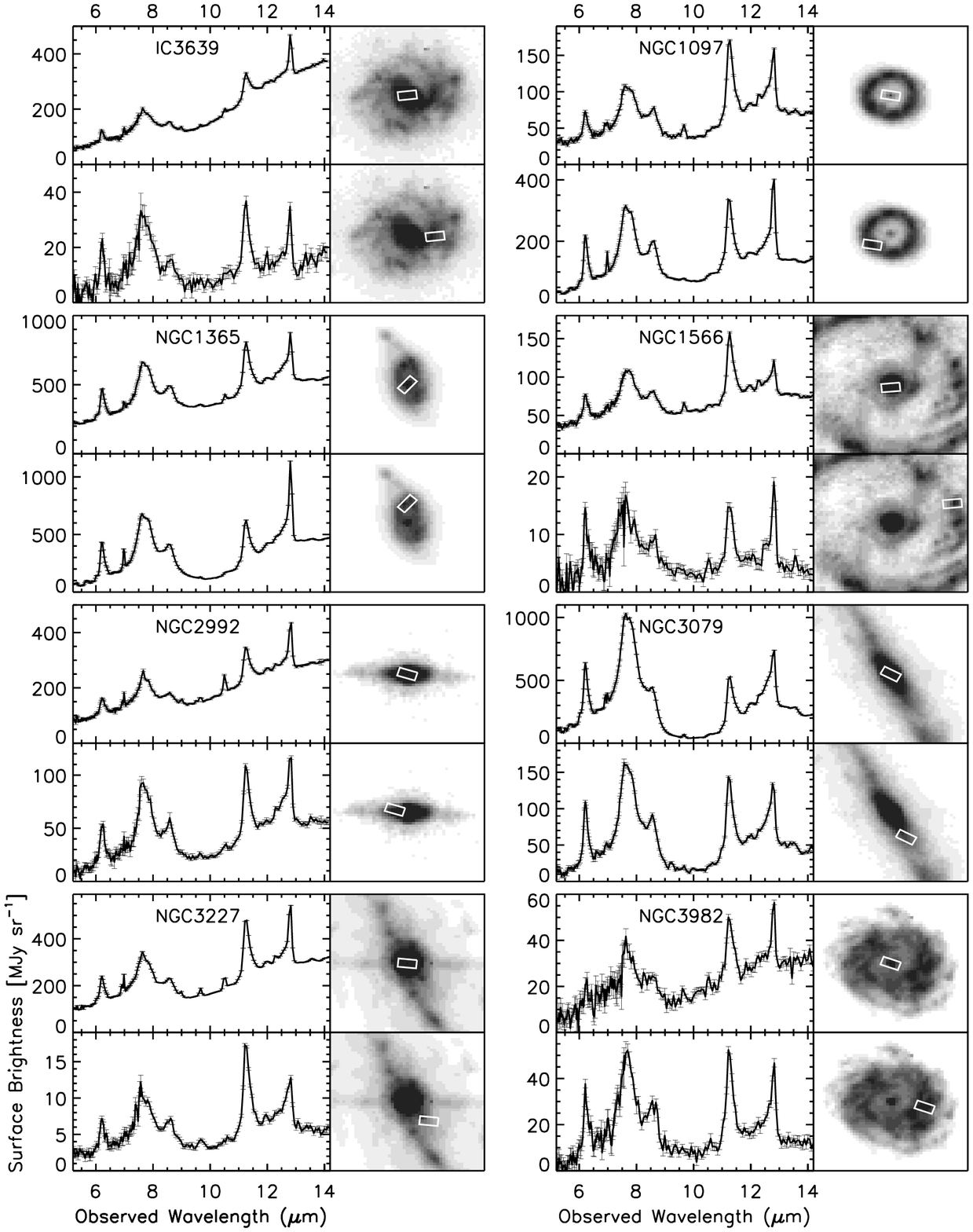}
\caption{Nuclear and off-nuclear spectra for the 21/35 Seyfert
  galaxies where off-nuclear regions were covered by the IRS slit and
  had sufficient S/N to detect the relevant aromatic features.  The
  panel to the right of each spectrum shows the corresponding
  $3.6\arcsec\times7.2\arcsec$ extraction region overlaid on the
  central $1\arcmin\times1\arcmin$ of an IRAC 8.0~$\mu$m image.}
\label{fig:off}
\end{center}
\end{figure*}

\begin{figure*}[!t]
\figurenum{\ref{fig:off}}
\begin{center}
\includegraphics[angle=0,scale=.9]{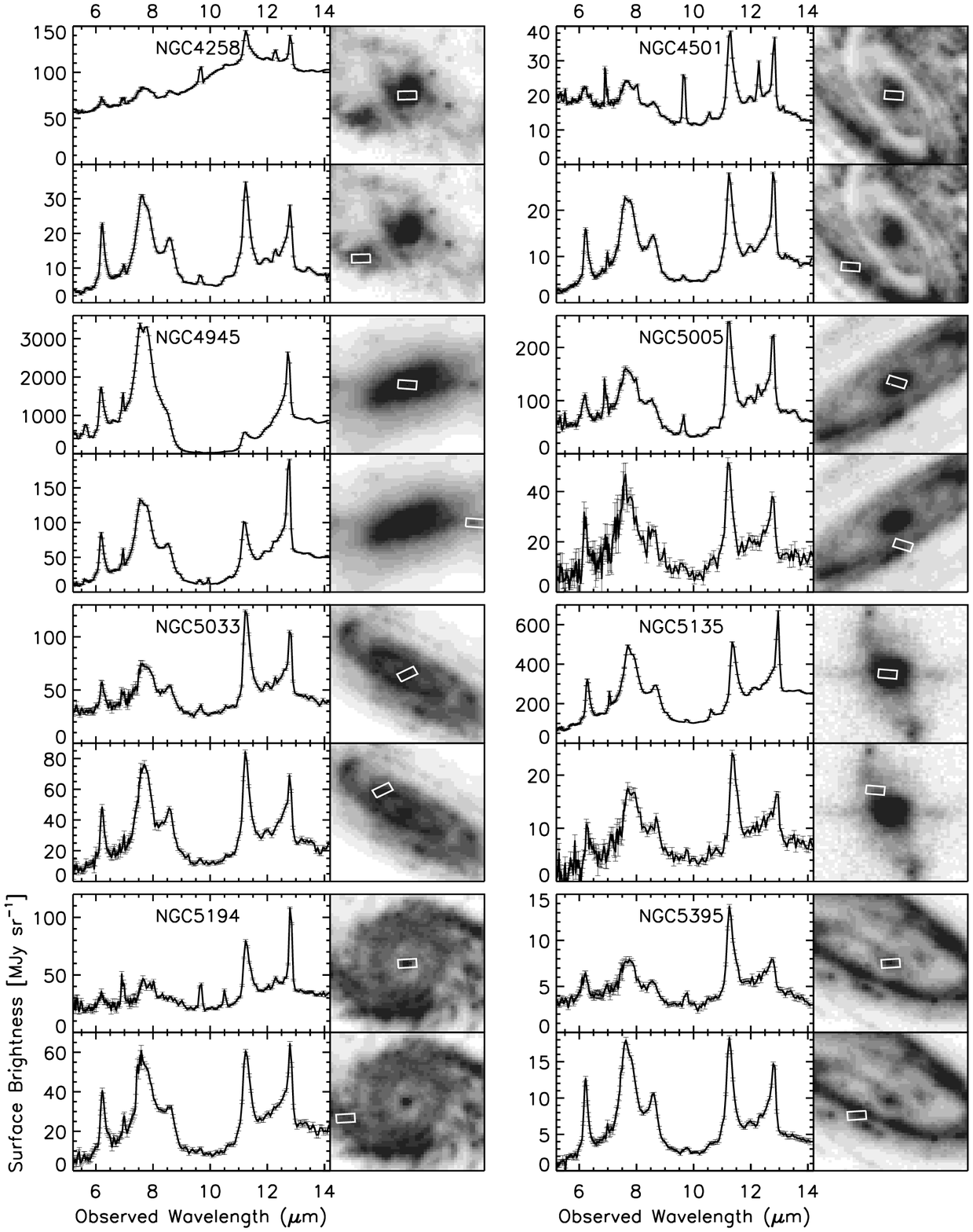}
\caption{{\it Continued}}
\label{fig:off_p2}
\end{center}
\end{figure*}

\begin{figure*}[!t]
\figurenum{\ref{fig:off}}
\begin{center}
\includegraphics[angle=0,scale=.9]{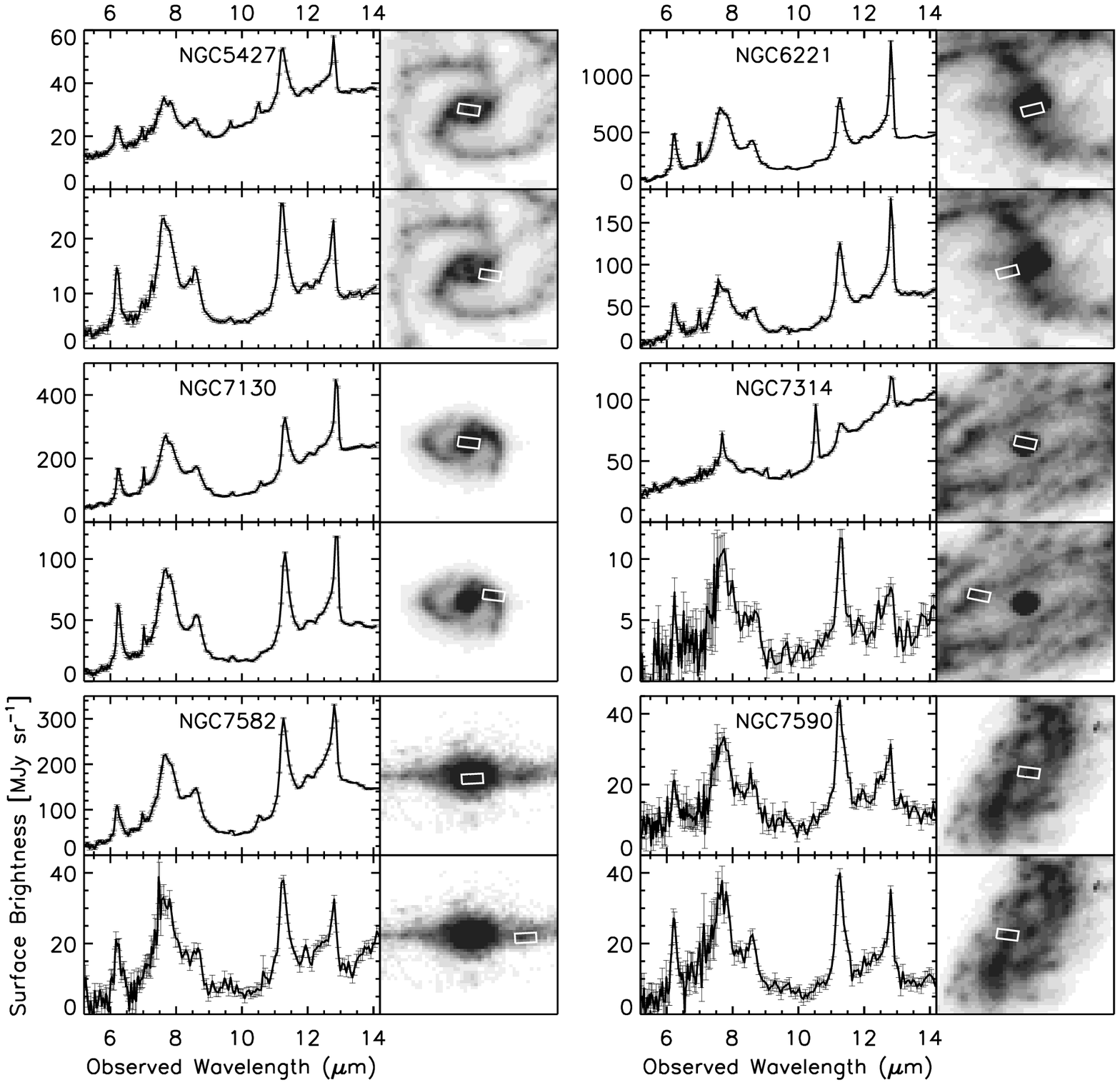}
\caption{{\it Continued}}
\label{fig:off_p3}
\end{center}
\end{figure*}

\clearpage

\begin{figure}[!t]
\begin{center}
\includegraphics[angle=0,scale=.43]{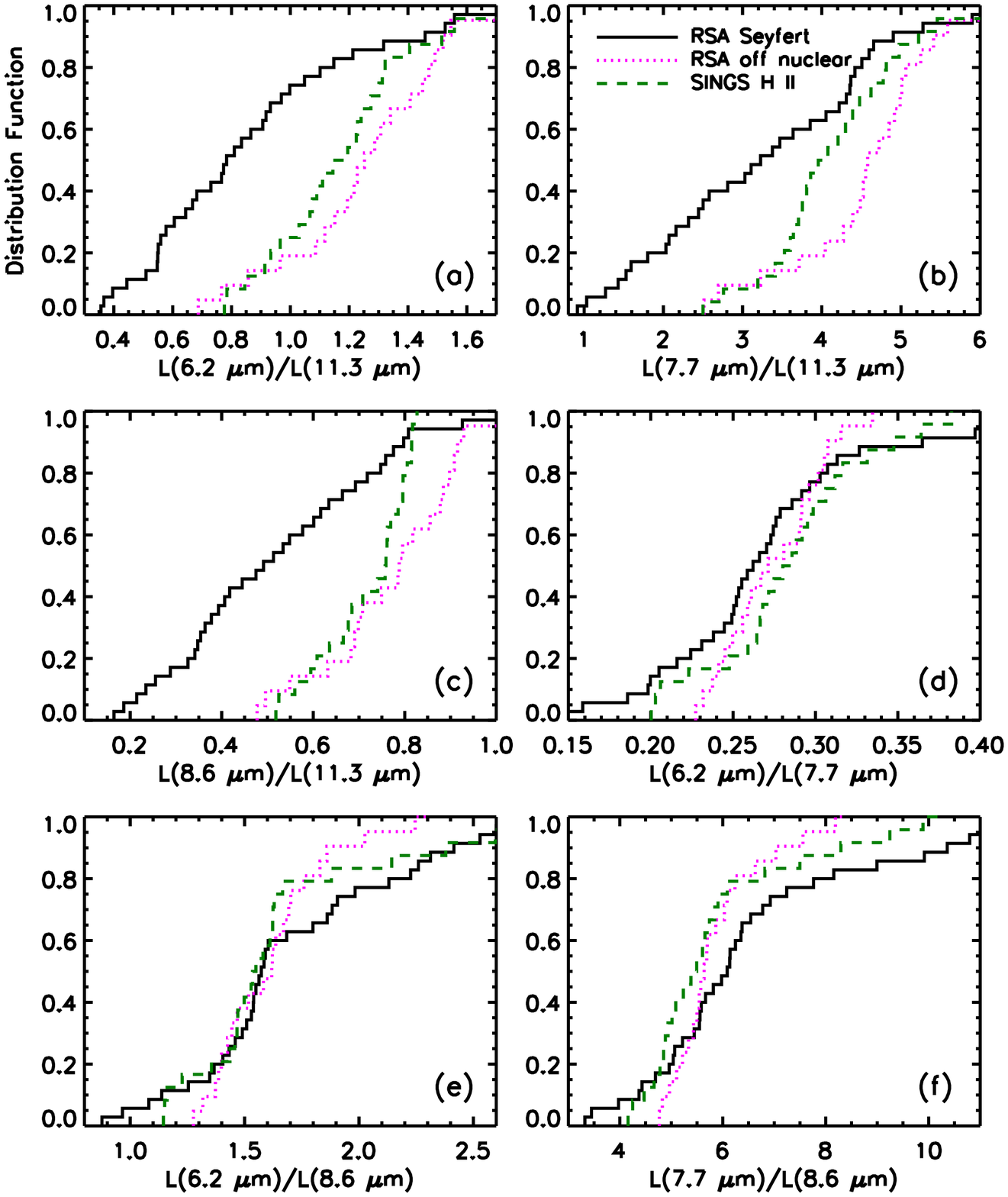}
\caption{The cumulative distribution of aromatic feature ratios for 35
  RSA Seyfert nuclei, 21 off-nuclear regions, and 27 SINGS
  \hii\ galaxies.  The first three panels show that the 6.2, 7.7, and
  8.6~$\mu$m features are systematically weaker relative to the
  11.3~$\mu$m feature for the Seyfert nuclei than for the off-nuclear
  regions or the SINGS \hii\ galaxies.  Panel (c), for example, shows
  that half of the RSA Seyfert nuclei have
  L(8.6~$\mu$m)/L(11.3~$\mu$m) ratios $<0.5$, whereas for the
  \hii\ galaxies, half have ratios $<0.75$.  The remaining three
  panels show that the ratios among the 6.2, 7.7, and 8.6~$\mu$m
  features show no significant differences between any of the
  samples.}
\label{fig:ratios}
\end{center}
\end{figure}

\begin{figure}[!t]
\begin{center}
\includegraphics[angle=90,scale=.36]{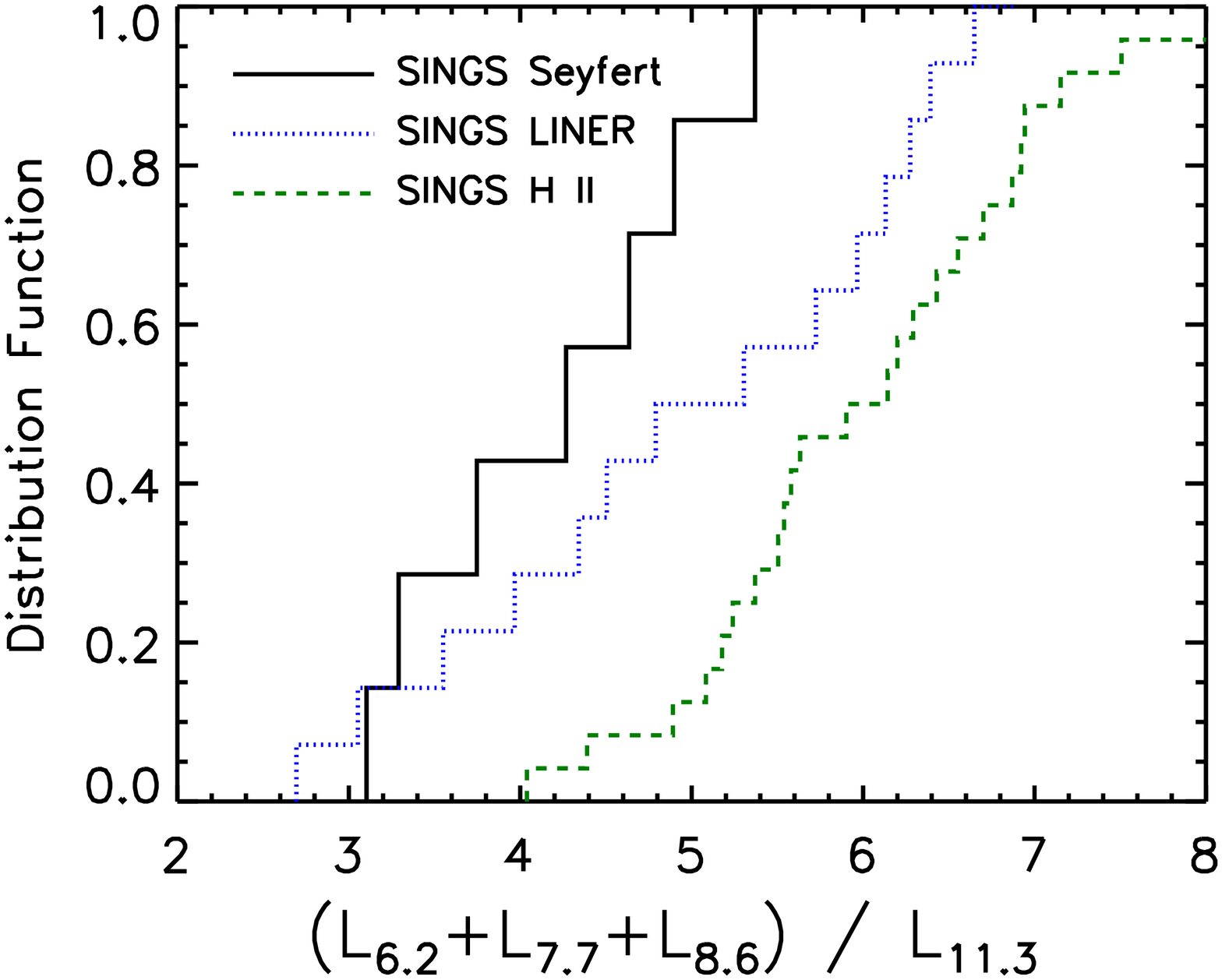}
\caption{The cumulative distribution of the ratio of 6.2, 7.7, and
  8.6~$\mu$m features to the 11.3~$\mu$m feature for SINGS galaxies
  with Seyfert, LINER, and \hii\ optical classifications.  This
  illustrates the result found by \citet{smi07a} that the Seyferts and
  LINERs have ratios that are significantly lower than the
  \hii\ galaxies.  The apparent difference between SINGS Seyferts and
  LINERs is not statistically significant.}
\label{fig:sings}
\end{center}
\end{figure}

\subsection{Aromatic Feature Ratio Distributions}

In Figure~\ref{fig:ratios}, we show the distribution of aromatic
feature ratios for the 35 RSA Seyfert nuclei and 21 off-nuclear
regions, as well as for 27/59 SINGS galaxies from \citet{smi07a} that
have \hii\ nuclear classifications (i.e., those that are not Seyferts
or LINERs).  We find that the L(6.2~$\mu$m)/L(11.3~$\mu$m),
L(7.7~$\mu$m)/L(11.3~$\mu$m), and L(8.6~$\mu$m)/L(11.3~$\mu$m) ratios
are systematically lower for the Seyfert nuclei than for the
off-nuclear regions or the SINGS \hii\ galaxies.  These differences
are all statistically significant with $p\leq0.003$ based on the
two-sample Kolmogorov--Smirnov (K-S) test (see Table~\ref{tab:stats}),
a non-parametric test that considers the maximum deviation between two
cumulative distribution functions \citep{pre92,wal03}.  On the other
hand, there are no significant differences between these feature
ratios for the off-nuclear regions and the SINGS \hii\ galaxies, so
the feature strengths in regions of star formation are consistent with
being drawn from the same parent distribution.  Furthermore, the
ratios among the 6.2, 7.7, and 8.6~$\mu$m features show no significant
differences between any of the samples.

\citet{smi07a} noted that the LINERs and Seyferts in the SINGS sample
were offset towards lower L(7.7~$\mu$m)/L(11.3~$\mu$m) ratios when
compared to the \hii\ galaxies.  We illustrate this result graphically
in Figure~\ref{fig:sings}, which shows the distribution functions of
the ratio of 6.2, 7.7, and 8.6~$\mu$m features to the 11.3~$\mu$m
feature for galaxies with Seyfert, LINER, and \hii\ optical
classifications.  Both Seyferts ($p=2\times10^{-3}$) and LINERs
($p=0.04$) have ratios that are significantly lower than the
\hii\ galaxies.  While the SINGS Seyferts have somewhat lower ratios
than the LINERs, this difference is not statistically significant, and
neither sample is statistically distinguishable from the RSA Seyferts.

\begin{figure}[!t]
\begin{center}
\includegraphics[angle=90,scale=.36]{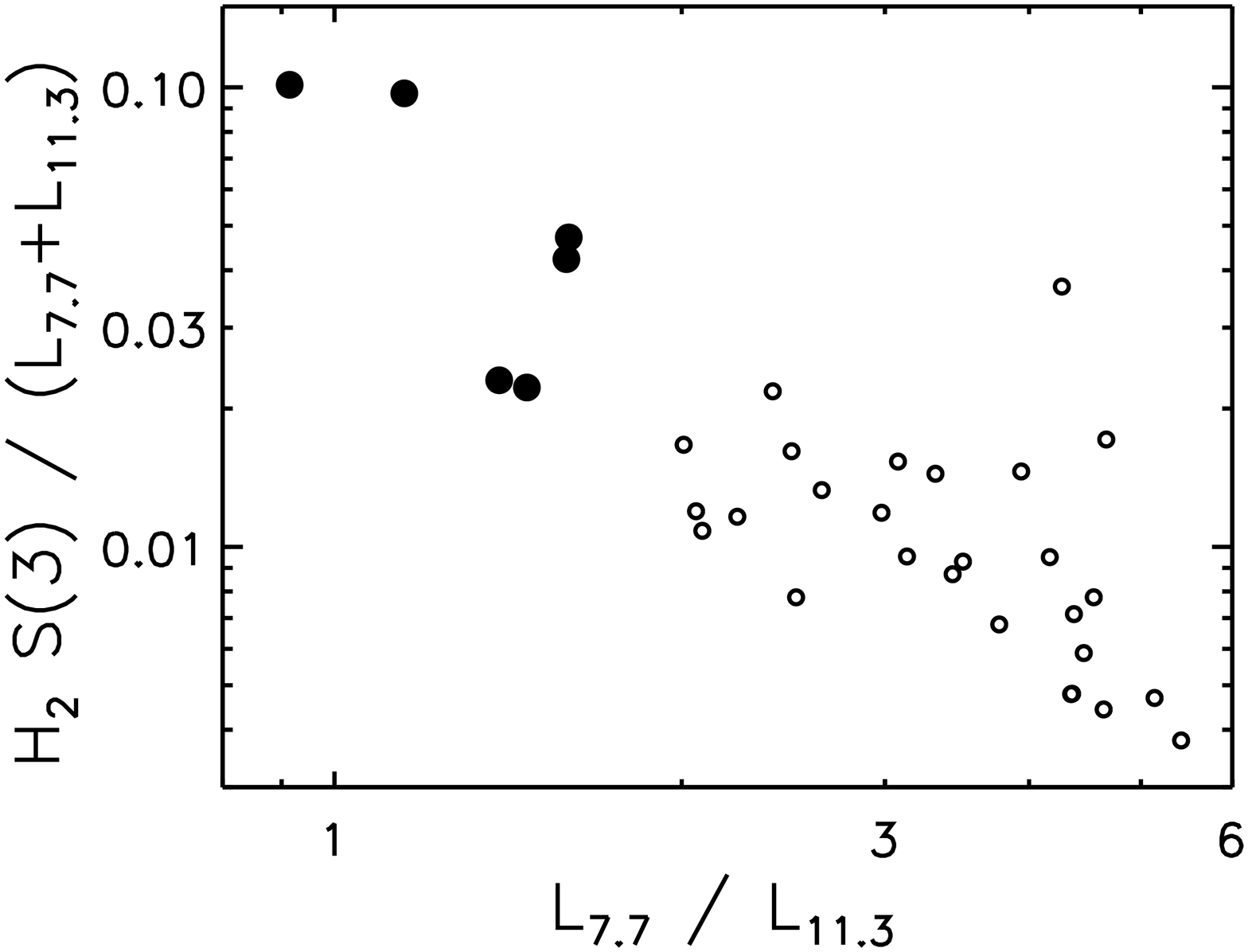}
\caption{The relationship between the strength of the H$_2$ S(3)
  rotational line, normalized to the strength of the aromatic
  features, and the L(7.7~$\mu$m)/L(11.3~$\mu$m) ratio for RSA Seyfert
  nuclei.  The sources with small L(7.7~$\mu$m)/L(11.3~$\mu$m) ratios
  also exhibit strong H$_2$ emission.  The most extreme sources with
  L(7.7~$\mu$m)/L(11.3~$\mu$m$)<1.6$ are highlighted with filled
  circles.}
\label{fig:h2}
\end{center}
\end{figure}

\begin{figure}[!t]
\begin{center}
\includegraphics[angle=0,scale=.43]{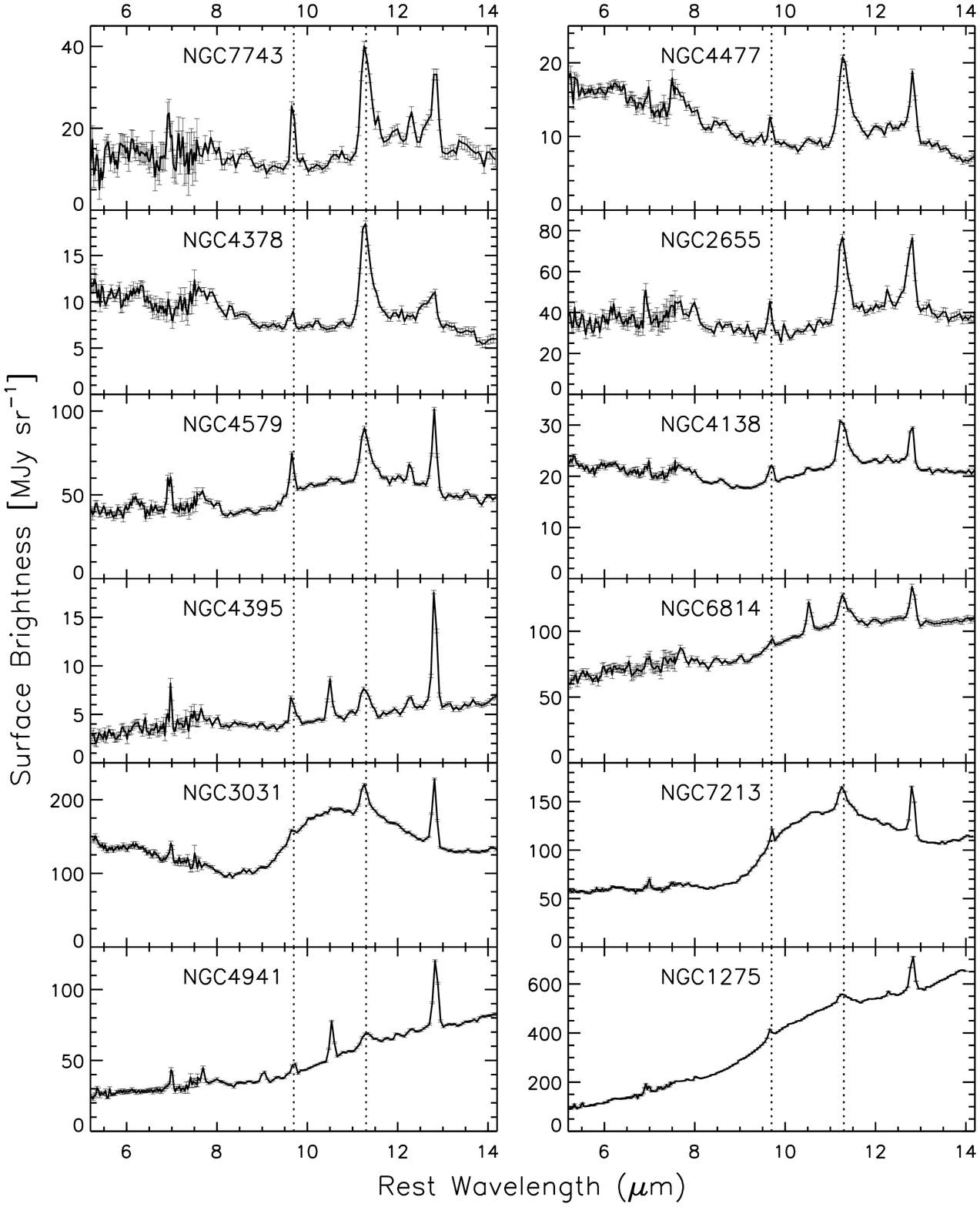}
\caption{Nuclear spectra for 12 additional RSA Seyfert nuclei that
  exhibit small L(7.7~$\mu$m)/L(11.3~$\mu$m) ratios and strong H$_2$
  S(3) lines, but were excluded from the sample due to a lack of 6.2,
  7.7, or 8.6~$\mu$m aromatic feature detections.  The spectra are
  sorted from top to bottom by the equivalent width of the 11.3~$\mu$m
  aromatic feature.  The wavelengths of the 9.67~$\mu$m H$_2$ S(3)
  line and the 11.3~$\mu$m aromatic feature are marked by dotted
  lines.}
\label{fig:cut}
\end{center}
\end{figure}

\subsection{Trends with H$_2$ emission}\label{results:h2}

Inspection of Figure~\ref{fig:spec} reveals that several Seyferts with
small L(7.7~$\mu$m)/L(11.3~$\mu$m) aromatic feature ratios also
exhibit a strong H$_2$ S(3) rotational line at 9.67~$\mu$m (e.g.,
NGC4501, NGC5194.)  To investigate this behavior, we plot the strength
of the H$_2$ S(3) line, normalized to the strength of the aromatic
features, as a function of the L(7.7~$\mu$m)/L(11.3~$\mu$m) ratio in
Figure~\ref{fig:h2}.  We find a strong anti-correlation in this plot
such that sources with the smallest L(7.7~$\mu$m)/L(11.3~$\mu$m)
ratios also have the strongest H$_2$ emission.  The Spearman's $\rho$
rank correlation coefficient is $-0.78$ with a probability
$p=3\times10^{-8}$ of no correlation, while Kendall's $\tau$ is
$-0.62$ with $p=1\times10^{-7}$; these non-parametric tests consider
the agreement between the ranks of quantities in pairs of measurements
\citep{pre92,wal03}, with coefficient values ranging from -1 (perfect
disagreement) to 1 (perfect agreement).  \citet{rou07} found that
H$_2$ rotational lines scale tightly with the aromatic features for
SINGS $\hii$ galaxies, but that Seyferts and LINERs often exhibit
excess H$_2$ emission, which they attribute to shocks.  We explore the
hypothesis that shocks cause both the excess H$_2$ emission and the
anomalous aromatic ratios for AGNs in Section~\ref{discussion:h2}.

Among the sources excluded from the above analysis due to a lack of
6.2, 7.7, or 8.6~$\mu$m aromatic feature detections, there are a
significant number with clearly detected 11.3~$\mu$m features and
H$_2$ S(3) lines.  In Figure~\ref{fig:cut}, we show the nuclear
spectra for a dozen of these sources, sorted by the equivalent width
of the 11.3~$\mu$m feature.  These spectra exhibit the small
L(7.7~$\mu$m)/L(11.3~$\mu$m) ratios and strong H$_2$ S(3) lines
characteristic of sources in the top-left of Figure~\ref{fig:h2}.  Due
to uncertainties associated with estimating the strength of weak,
broad features and determining robust upper limits (e.g., proper
continuum placement), we do not include any of these sources in our
subsequent analysis.  However, their behavior is consistent with that
in Figure~\ref{fig:h2} and supports the reality of the trend between
aromatic feature characteristics and H$_2$ line strength.

\subsection{Evidence for extinction of aromatic features}\label{results:si}

The sources with the largest L(7.7~$\mu$m)/L(11.3~$\mu$m) ratios,
NGC4945 and NGC3079, also have the strongest silicate absorption
features.  This suggests that the 11.3~$\mu$m feature is being
significantly attenuated, consistent with previous results for
starburst and luminous infrared galaxies \citep[e.g.,][]{bra06,per10},
and implies that a significant fraction of the silicate-absorbing
material is extended relative to the regions that produce the aromatic
features.  Although the aromatic feature measurements in PAHFIT are
corrected for extinction, in cases as extreme as these two galaxies
the resulting feature strengths are highly uncertain.  For all other
galaxies in our sample, the inferred extinctions are $<50\%$ for all
features.

\vspace{1cm}

\section{Discussion}

The result that Seyfert galaxies exhibit weak 6.2, 7.7, and 8.6~$\mu$m
aromatic features relative to the 11.3~$\mu$m feature could be
explained by radiative or mechanical processing of the molecular
carriers by the active nucleus.  Here we explore the relevant physical
and chemical effects that could modify the observed feature strengths.

\subsection{Ionization Balance}

Previous experimental \citep[e.g.,][]{szc93,hud95} and theoretical
\citep[e.g.,][]{def93,lan96} work on PAHs has shown that the C--C
stretching modes that produce the 6.2 and 7.7~$\mu$m features, as well
as the C--H in-plane bending modes that produce the 8.6~$\mu$m
feature, are more efficiently excited in ionized molecules.  The
ratios of these features to the 11.3~$\mu$m feature, which is produced
by C--H out-of-plane bending modes, are lower for neutral molecules
\citep[see Figure 1 of][]{all99}.  The fraction of ionized aromatic
molecules is set by the balance between ionization and recombination,
which depends on the UV radiation field density ($G_0$), the gas
temperature ($T)$, and the electron density ($n_e$) according to
$G_0~T^{1/2}/n_e$ \citep{bak94}.

\citet{gal08} argued that the variations in aromatic feature ratios
for a heterogeneous sample of 50 objects (including Galactic regions,
Magellanic \hii\ regions, and galaxies, as well as spatially resolved
regions within seven of those objects) are controlled by this
ionization balance.  Similar to our results, they found that the
relative strengths of 6.2, 7.7, and 8.6~$\mu$m features showed little
variation, while the ratios between these features and the 11.3~$\mu$m
feature varied by an order of magnitude.  This hypothesis is also
supported by observations of Galactic reflection nebulae by
\citet{job96} and \citet{bre05}, who found decreasing
L(8.6~$\mu$m)/L(11.3~$\mu$m) and L(7.7~$\mu$m)/L(11.3~$\mu$m) ratios
as a function of distance from the ionizing source, consistent with an
increasing neutral fraction.

\begin{figure}[!t]
\begin{center}
\includegraphics[angle=90,scale=.35]{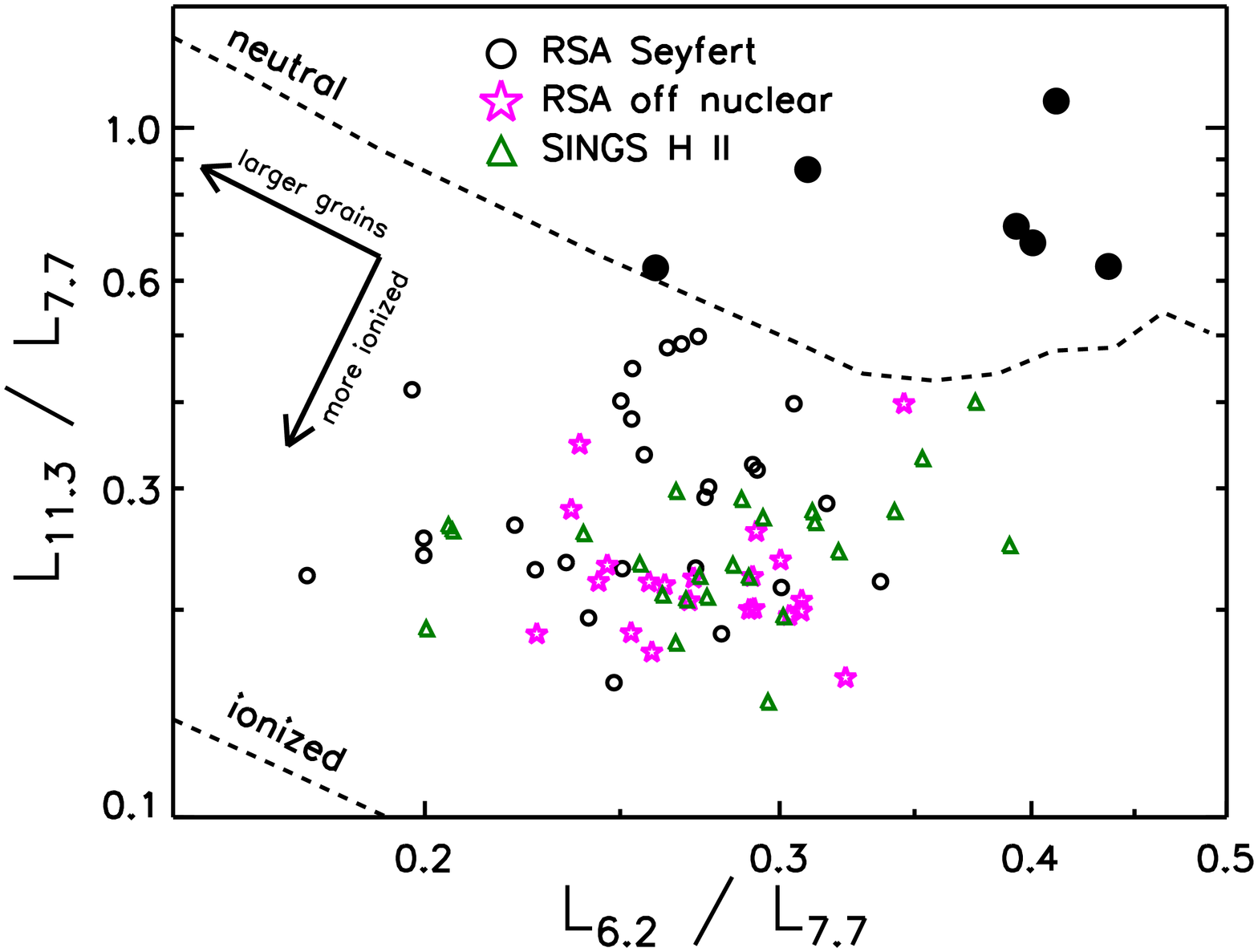}
\caption{The relative strengths of the 6.2, 7.7, and 11.3~$\mu$m
  features for RSA Seyfert nuclei, off-nuclear regions, and SINGS
  \hii\ galaxies compared with model predictions from \citet{dra01}
  for neutral and ionized PAHs.  The dashed lines correspond to
  predictions for completely neutral and completely ionized molecules;
  the permitted region of the diagram is bounded by these two lines.
  The arrows illustrate the effects of increasing grain size and
  increasing ionization on the aromatic feature ratios.  The Seyferts
  highlighted as filled circles in Figure~\ref{fig:h2} all lie beyond
  the range of model predictions, even for completely neutral
  molecules.}
\label{fig:draine}
\end{center}
\end{figure}

To compare with model expectations for ionized and neutral aromatic
molecules, we plot L(11.3~$\mu$m)/L(7.7~$\mu$m)
v. L(6.2~$\mu$m)/L(7.7~$\mu$m) for the Seyfert nuclei, off-nuclear
sources, and SINGS \hii\ galaxies in Figure~\ref{fig:draine}.  This
can be compared with Figure~16 of \citet{dra01} and Figure~5 of
\citet{odo09}, although we use a condensed plot range.  We find that a
number of Seyferts lie beyond the range of model predictions, even for
completely neutral aromatic molecules; these are the 6/35 Seyferts
with L(11.3~$\mu$m)/L(7.7~$\mu$m$)>0.6$: NGC5194, NGC4501, NGC4639,
NGC1433, NGC2639, and NGC5005.  While such
L(11.3~$\mu$m)/L(7.7~$\mu$m) ratios could be produced by large ($>200$
C atoms) neutral molecules, they would be expected to have
L(6.2~$\mu$m)/L(7.7~$\mu$m$)<0.25$, which is inconsistent with the
data.  Similar extreme aromatic band strengths were observed by
\citet{rea00} for the quiescent molecular cloud SMC B1 No. 1, and
\citet{li02} were unable to reproduce the observed band ratios even
with completely neutral grains.

The above comparison is for a single Milky Way--based model, and
laboratory studies have found larger L(11.3~$\mu$m)/L(7.7~$\mu$m)
ratios for neutral PAHs, but it does illustrate the difficulty in
explaining our results for Seyfert galaxies in terms of a low ionized
fraction.  Furthermore, under the assumption that aromatic features
are produced by star formation (see Section~\ref{sec:excite}), the
temperatures and densities of the aromatic-emitting regions should be
typical of PDRs, whereas the UV radiation field would likely be
enhanced by the AGN.  This implies that the ionized fraction would be
higher, not lower.  Thus, ionization balance arguments appear unable
to explain the behavior of the aromatic features around AGNs.

\subsection{Grain Size}

Smaller aromatic molecules contribute preferentially to the
shorter-wavelength features \citep[e.g.,][]{sch93}, but they are
subject to photodestruction by the UV radiation field and collisional
destruction by shocks.  Based on laboratory studies, \citet{joc94}
found a critical size of 30--40 C atoms, below which PAHs would mainly
be photodissociated, while \citet{all96} suggested a larger critical
value of 50 C atoms based on their models.  \citet{lep03} agreed that
small PAHs with $15$--20 C atoms or fewer would be destroyed in most
environments, but their models indicated that PAHs in the 20--30 C
atom range may survive, albeit with most of their peripheral H atoms
stripped away, while larger PAHs would survive with their H atoms
intact.  \citet{mic10a} found that PAHs with 50 C atoms would not
survive in shocks with velocities greater than 100~km~s$^{-1}$, while
PAHs with 200 C atoms would be destroyed by shocks with velocities
above 125~km~s$^{-1}$.

Destruction of the smallest molecules is expected to result in the 6.2
and 7.7~$\mu$m features being suppressed relative to the 11.3~$\mu$m
feature, as well as the 6.2~$\mu$m feature being suppressed relative
to the 7.7~$\mu$m feature \citep[e.g.,][]{dra01,gal08}.  The former
effect is clearly seen in Figure~\ref{fig:ratios}, but the latter is
not.  Thus, the hypothesis that small-grain destruction can explain
the observed ratios is only tenable if the molecules that produce the
6.2, 7.7, and 8.6~$\mu$m features are destroyed with similar
efficiency, which is inconsistent with existing models.

\subsection{Hydrogenation and Molecular Structure}

The level of hydrogenation of the aromatic molecules will affect the
number of C--H bonds and therefore the relative strength of the C--H
and C--C vibrational modes.  An increase in the C--H/C--C ratio was
proposed by \citet{rea00} to explain the large
L(11.3~$\mu$m)/L(7.7~$\mu$m) ratio observed in SMC B1 No. 1, although
\citet{dra01} and \citet{li02} point out that PAHs with $>30$ C atoms
are already expected to be fully hydrogenated.  Some range in
C--H/C--C ratios, even for fully hydrogenated molecules, is
facilitated by the structure of the C skeleton, which can be compact
with more C--C bonds or open with more C--H bonds \citep[e.g.,
  pericondensed PAHs v. catacondensed PAHs,][]{tie05}.  The structure
also affects the number of adjacent C--H groups per aromatic ring, and
therefore the relative strengths of the 11.3~$\mu$m feature, which is
produced by solo C--H bonds, and the 12.7~$\mu$m feature, which is
produced by C--H multiplets \citep[e.g.,][]{hon01}.  For example,
based on the large L(12.7~$\mu$m)/L(11.3~$\mu$m) ratio for SMC B1
No. 1, \citet{ver02} argued for a compact structure with a higher
incidence of C--H multiplets.

\begin{figure}[!t]
\begin{center}
\includegraphics[angle=90,scale=.36]{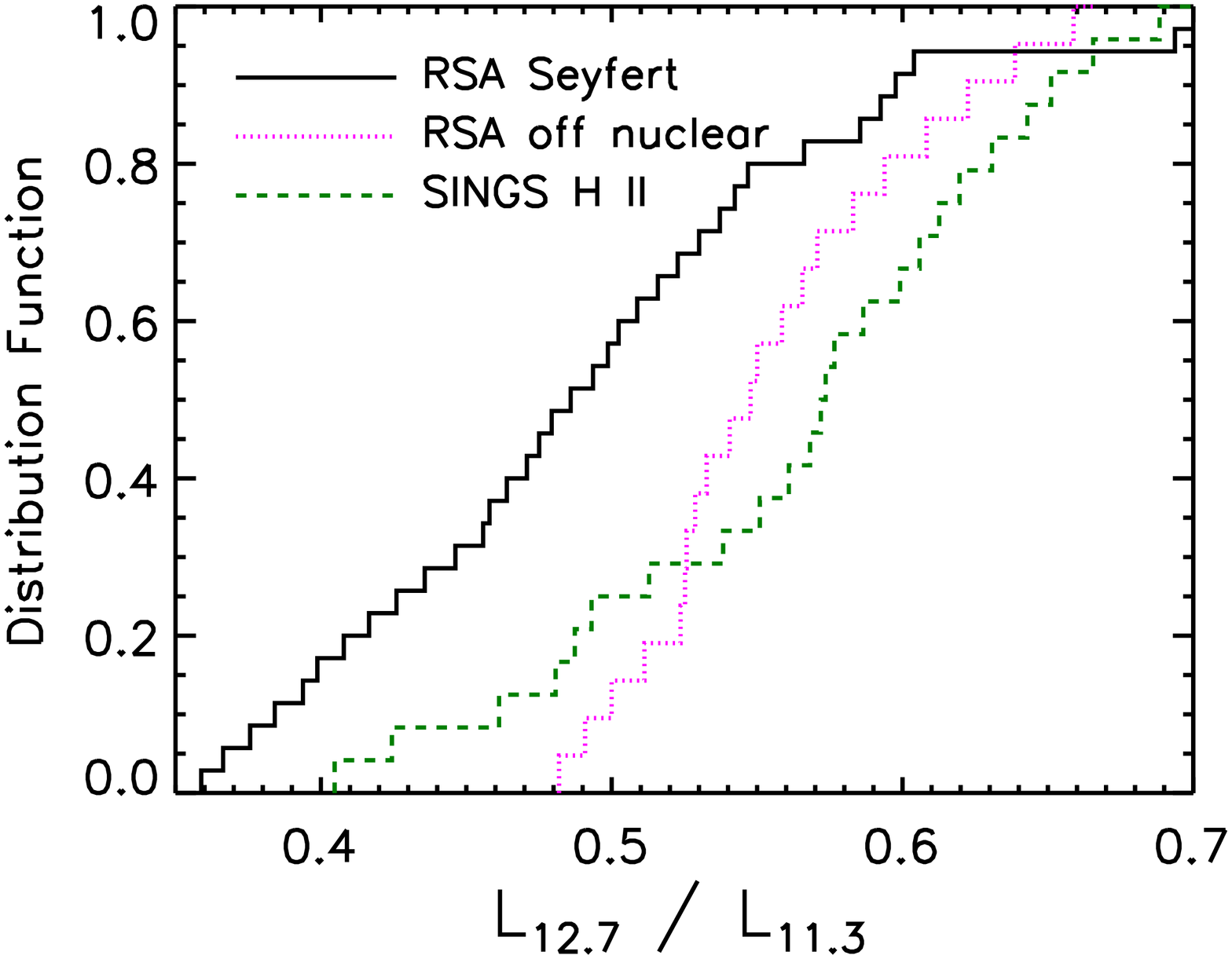}
\caption{The cumulative distribution of L(12.7~$\mu$m)/L(11.3~$\mu$m)
  ratios for RSA Seyfert nuclei, off-nuclear regions, and SINGS
  \hii\ galaxies.  The result that Seyfert nuclei exhibit
  significantly smaller ratios suggests aromatic molecules that have
  fewer adjacent C--H groups.}
\label{fig:ch}
\end{center}
\end{figure}

To investigate such behavior, we plot the
L(12.7~$\mu$m)/L(11.3~$\mu$m) ratios for Seyfert nuclei, off-nuclear
regions, and SINGS \hii\ galaxies in Figure~\ref{fig:ch}.  The Seyfert
nuclei exhibit significantly smaller ratios ($p\leq0.001$), while the
ratios for off-nuclear regions and SINGS \hii\ galaxies are not
distinguishable.  This implies that the aromatic molecules in Seyfert
nuclei may have fewer C--H multiplets.  Thus a scenario where AGN
processing or environment results in open, uneven molecular structures
with higher C--H/C--C ratios and fewer adjacent C--H groups could
qualitatively explain the observed 6--13~$\mu$m aromatic spectra.

\subsection{The role of AGN-driven shocks}\label{discussion:h2}

As presented in Section~\ref{results:h2} and Figure~\ref{fig:h2}, the
Seyfert galaxies with the smallest L(7.7~$\mu$m)/L(11.3~$\mu$m)
aromatic feature ratios also exhibit the strongest H$_2$ S(3)
emission, which probes hot molecular gas (upper level temperature
2500~K).  The incidence of this excess H$_2$ emission does not scale
with AGN luminosity, indicating that shock excitation is more
important than X-ray heating \citep[e.g.,][]{rou07}.  A connection
between shock-heated, H$_2$-emitting gas and small
L(7.7~$\mu$m)/L(11.3~$\mu$m) ratios was found by \citet{ogl07} for the
radio galaxy 3C 326 and by \citet{gui10} for Stephan's Quintet, a
compact group of interacting galaxies exhibiting a large-scale shock
\citep[e.g.,][]{app06,clu10}.  Similarly, \citet{kan08} found strong
H$_2$ emission and small L(7.7~$\mu$m)/L(11.3~$\mu$m) ratios in a
sample of local elliptical galaxies, many of which host low-luminosity
AGNs.  More recently, \citet{veg10} affirmed this result for a sample
of four early-type galaxies classified as LINERs, and they argued that
shock processing of aromatic molecules may be responsible for the
observed behavior.  As discussed above, the Seyferts and LINERs in the
SINGS sample also exhibit smaller L(7.7~$\mu$m)/L(11.3~$\mu$m) ratios
\citep{smi07a} and stronger H$_{2}$ emission \citep{rou07} than do the
\hii\ galaxies.

\begin{figure}[!t]
\begin{center}
\includegraphics[angle=0,scale=.44]{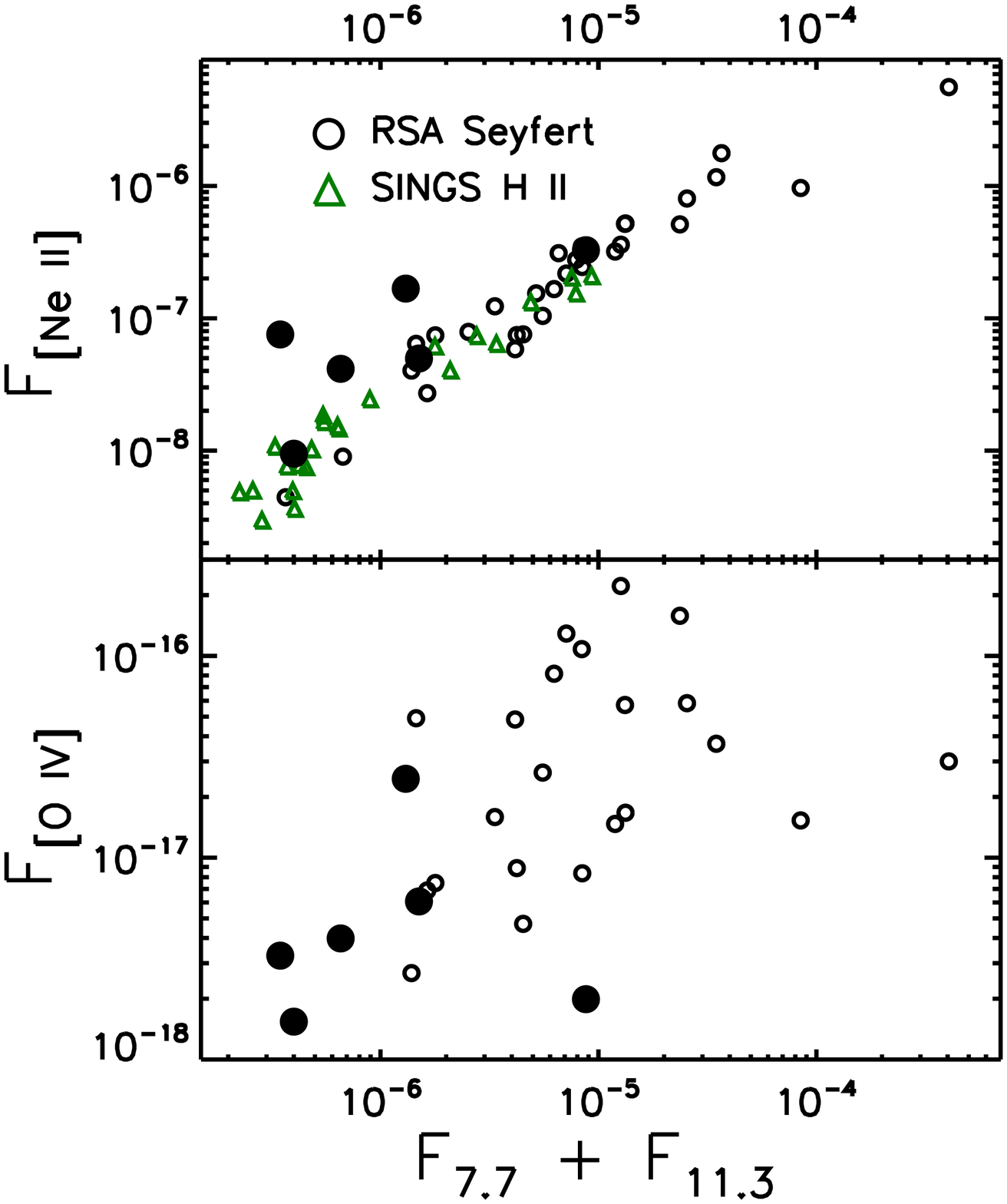}
\caption{The relationship between the aromatic features and the
  \neii\ and \oiv\ emission lines.  The filled circles correspond to
  the RSA Seyferts defined in Figure~\ref{fig:h2} that have the
  smallest L(7.7~$\mu$m)/L(11.3~$\mu$m) ratios.  The strong
  correlation with \neii, which traces star formation, and the weak
  correlation with \oiv, which traces AGN activity, implies that the
  aromatic features are primarily associated with star formation.
  Most of the Seyfert nuclei lie on the relationship between aromatic
  feature and \neii\ emission for \hii\ galaxies; the only outliers
  are among the sources highlighted with filled circles, which have
  extreme aromatic feature ratios (see Figures~\ref{fig:h2} and
  \ref{fig:draine}).  The aromatic feature and \neii\ emission values
  are in surface brightness units (W~m$^{-2}$~sr$^{-1}$), while the
  \oiv\ values, taken from \citet{dia09}, are in flux units
  (W~m$^{-2}$).}.
\label{fig:ew2}
\end{center}
\end{figure}

Shocks are expected to have profound impacts on interstellar dust via
shattering in grain-grain collisions and sputtering in ion-grain
collisions \citep[e.g.,][]{jon94,jon96}.  Aromatic features are
nonetheless observed in the shocked environments of supernova remnants
\citep[e.g.,][]{tap06,rea06} and galactic winds
\citep[e.g.,][]{tac05,eng06}.  The observed emission may come from
entrained clumps that are not fully exposed to the shock or the hot,
post-shock gas \citep{mic10a,mic10b}.  \citet{mic10a} study the
processing of small carbon grains ($N_C\leq200$, corresponding to
aromatic molecules) by interstellar shocks and find that their
molecular structure is severely denatured for shock velocities of
75--100~km~s$^{-1}$ and they are completely destroyed when
$v\geq125$~km~s$^{-1}$.  The effect of this shock processing on the
observed aromatic feature ratios is not known.  A possibility that
could explain the association of modified aromatic feature ratios with
strong H$_2$ emission is that shocks may leave open, uneven structures
in the surviving aromatic molecules.  We note that AGN-driven shocks,
if responsible for the observed behavior, do not strongly suppress the
11.3~$\mu$m feature (see Section~\ref{sec:sfr}).

\begin{figure}[!t]
\begin{center}
\includegraphics[angle=90,scale=.36]{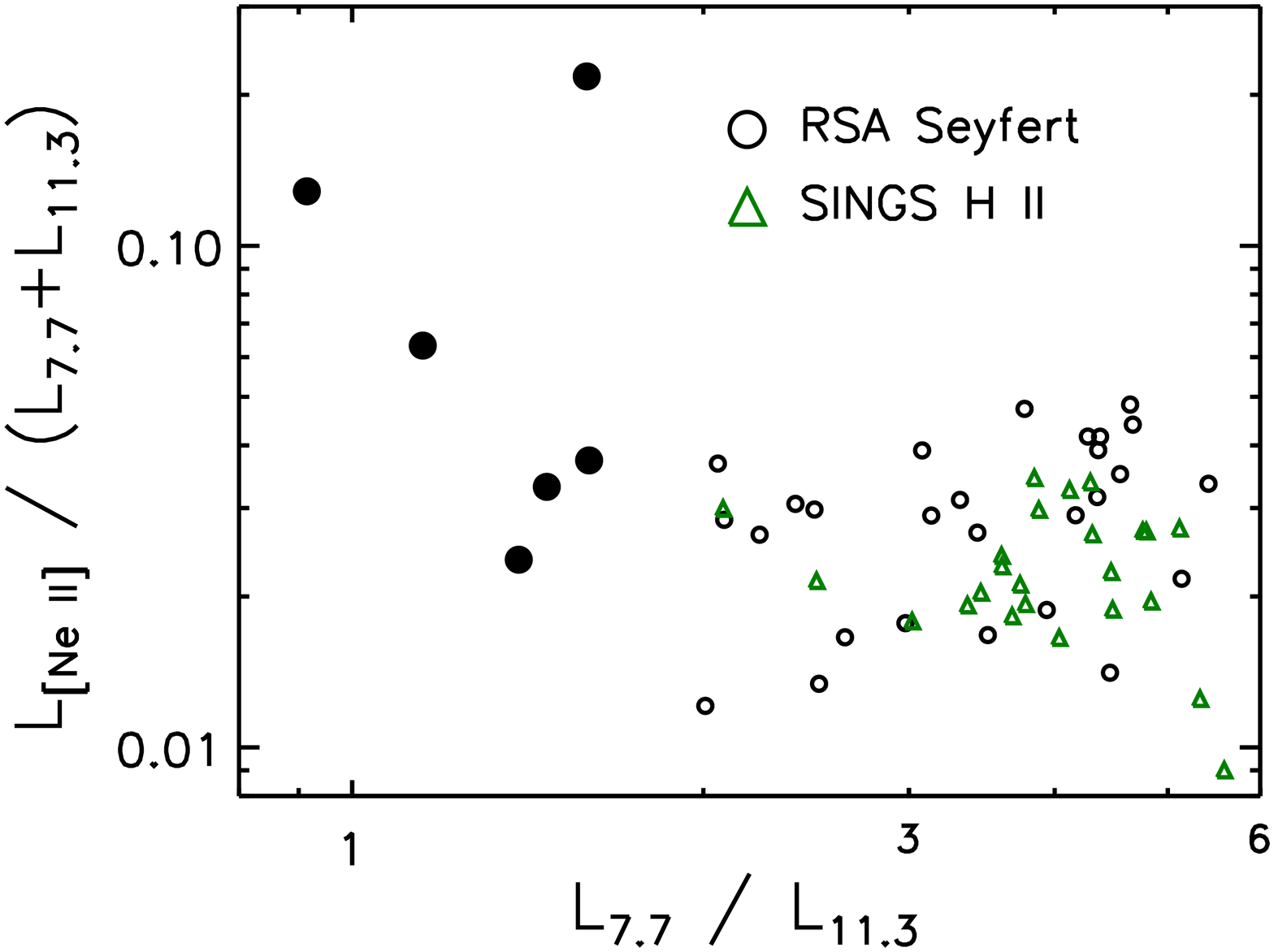}
\caption{The relationship between aromatic feature and \neii\ emission
  as a function of the L(7.7~$\mu$m)/L(11.3~$\mu$m) ratio.  The filled
  circles correspond to the RSA Seyferts defined in
  Figure~\ref{fig:h2} that have the smallest
  L(7.7~$\mu$m)/L(11.3~$\mu$m) ratios.  This confirms that the sources
  with suppressed aromatic features, relative to \neii, have the
  smallest L(7.7~$\mu$m)/L(11.3~$\mu$m) ratios.}
\label{fig:ew3}
\end{center}
\end{figure}

\begin{figure}[!t]
\begin{center}
\includegraphics[angle=0,scale=.44]{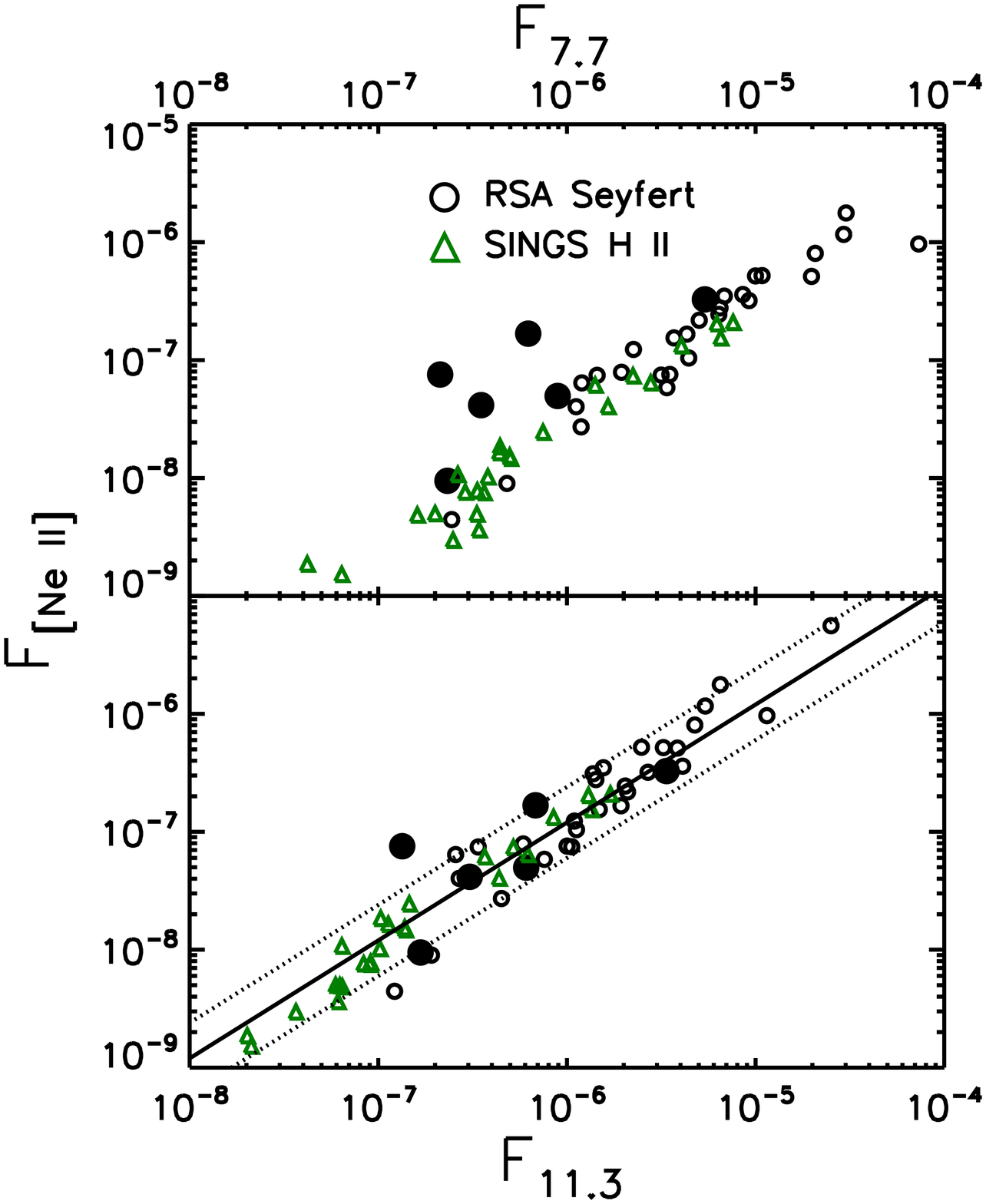}
\caption{The relationship between \neii\ emission and the 7.7 and
  11.3~$\mu$m aromatic features.  The filled circles correspond to the
  RSA Seyferts defined in Figure~\ref{fig:h2} that have the smallest
  L(7.7~$\mu$m)/L(11.3~$\mu$m) ratios.  While the 7.7~$\mu$m feature
  can be strongly suppressed, the 11.3~$\mu$m feature is still a
  robust tracer of the SFR.  The solid line in the bottom panel
  corresponds to the median ratio
  L$_{\scriptsize{\neii}}$/L$_{11.3}=0.12$, and the dotted lines
  correspond to factors of two above and below this median value.
  Scatter in this ratio is expected because \neii\ traces somewhat
  younger stellar populations than do the aromatic features.  All
  values are in surface brightness units (W~m$^{-2}$~sr$^{-1}$).}
\label{fig:ew2a}
\end{center}
\end{figure}

\subsection{Could the aromatic features be excited by the AGN?}\label{sec:excite}

\citet{smi07a} speculated that the AGNs could directly excite aromatic
emission.  If this were the case, SFRs estimated from aromatic
features would be overestimated due to this AGN contribution.  To
investigate the relationship between star formation rate, AGN
luminosity, and aromatic feature strength, we plot the fluxes of the
\neii\ and \oiv\ emission lines versus those of the 7.7~$\mu$m and
11.3~$\mu$m aromatic features in Figure~\ref{fig:ew2}.  The
\neii\ line has an ionization potential of 21~eV and is a reasonable
tracer of the SFR \citep[e.g.,][]{ho07}, while the \oiv\ line has an
ionization potential of 55~eV and traces the AGN intrinsic luminosity
\citep[e.g.,][]{mel08,dia09,rig09}.  Figure~\ref{fig:ew2} shows the
strong correlation between \neii\ and aromatic feature strength for
RSA Seyferts (Spearman's $\rho=0.93$), which matches the relationship
for SINGS \hii\ galaxies, and it shows the weak correspondence between
\oiv\ and aromatic feature strength (Spearman's $\rho=0.39$).  This
confirms that the aromatic features are primarily tracing
star-formation activity.

The Seyferts that are outliers in the \neii--aromatic feature
relationship have weak aromatic features, and we show in
Figure~\ref{fig:ew3} that these correspond to the sources with the
smallest L(7.7~$\mu$m)/L(11.3~$\mu$m) ratios.  There are no examples
with stronger aromatic features as might be expected if the AGN were
exciting additional emission.  The three obvious outliers, NGC2639,
NGC4501, and NGC5194, all have $\oiv/\neii<0.25$, implying that the
AGN contribution to \neii\ is $<10$\% \citep[e.g.,][]{stu02}.  We note
that the incidence of modified aromatic spectra does not show a
dependence on AGN luminosity, confirming the results of \citet{bau10},
who found no correlation between the L(6.2~$\mu$m)/L(11.3~$\mu$m)
ratio and \nev\ luminosity.

\subsection{Use of Aromatic Features to Determine SFRs}\label{sec:sfr}

Several studies \citep[e.g.,][]{sch06, lut08, shi09} have used the 6.2
and 7.7~$\mu$m aromatic features to measure the SFRs in AGN host
galaxies.  The result that some AGNs exhibit suppressed
short-wavelength aromatic features (e.g., the outliers in
Figure~\ref{fig:ew3}) suggests that such SFR measurements may be
underestimated.  To determine whether the 11.3~$\mu$m feature is
robust to such effects, we plot separately the relationships between
\neii\ and the 7.7 and 11.3~$\mu$m features in Figure~\ref{fig:ew2a}.
We find that almost all of the RSA Seyferts, including those with
anomalously high L$_{\scriptsize{\neii}}$/L$_{7.7}$ values, are within
a factor of two of the median value
L$_{\scriptsize{\neii}}$/L$_{11.3}=0.12$.  Scatter in this ratio is
expected as a function of the age of the stellar population because
21~eV photons from young stars ($<10$~Myr) are required to produce
\neii, while somewhat older stars can produce 6--13.6~eV UV photons
that excite aromatic emission \citep[e.g.,][]{pee04,dia10,per10}.
Silicate absorption will tend to increase the observed ratio, but this
is only a significant effect for sources like NGC4945 and NGC3079 (see
Section~\ref{results:si}).  While SFR estimates based on the
11.3~$\mu$m feature are still subject to the uncertainties that apply
to \hii\ galaxies \citep[e.g.,][]{smi07a}, such measurements for AGN
hosts appear to be robust to the effects of AGN- and shock-processing
of aromatic molecules.

\section{Conclusions}

We have shown that the relative strengths of the mid-IR aromatic
features for Seyfert galaxies differ significantly from those for
star-forming galaxies, with the 6.2, 7.7, and 8.6~$\mu$m features
being suppressed relative to the 11.3~$\mu$m feature in Seyferts.  The
sources with the smallest L(7.7~$\mu$m)/L(11.3~$\mu$m) aromatic
feature ratios also exhibit the strongest H$_2$ S(3) rotational lines,
which likely trace shocked gas (see Figure~\ref{fig:h2}).  We explore
the relevant physical and chemical effects that could produce the
observed aromatic spectra.  An enhanced fraction of neutral aromatic
molecules could produce qualitatively similar behavior, but the
observed ratios lie beyond model predictions for completely neutral
molecules and the presence of an AGN would be expected to increase the
level of ionization rather than reduce it.  Destruction of the
smallest aromatic molecules could explain the suppression of shorter
wavelength features, but the expected variations in the relative
strengths of the 6.2, 7.7, and 8.6~$\mu$m features are not seen.  A
modification of the molecular structure that enhances the C--H/C--C
ratio could reproduce the observed behavior, and an open C skeleton
with fewer adjacent C--H groups would furthermore explain the reduced
strength of the 12.7~$\mu$m feature.  Given the connection between
strong H$_2$ emission and modified aromatic ratios, we speculate that
shock processing could produce such structures.  Finally, we show that
the aromatic features correlate well with \neii\ (i.e., star
formation) but not with \oiv\ (i.e., AGN luminosity), indicating that
AGN excitation of aromatic emission is not significant and that
aromatic-based estimates of the SFR are generally reasonable.  There
are a few outliers with strong H$_2$ emission, small
L(7.7~$\mu$m)/L(11.3~$\mu$m) ratios, and small aromatic/\neii\ ratios,
but for these sources the 11.3~$\mu$m feature is still a reasonably
robust tracer of the SFR.

\acknowledgments

We acknowledge useful discussions with and assistance from Anthony
Jones, Yong Shi, Amelia Stutz, Alexander Tielens, Jonathan Trump, and
Gregory Walth.  We thank the anonymous referee for helpful suggestions
that have improved the manuscript.  This work was supported by
contract 1255094 from Caltech/JPL to the University of Arizona.

{\it Facilities:} \facility{Spitzer}

\clearpage


\begin{deluxetable}{lccccccc}
\tabletypesize{\scriptsize}
\tablecaption{Nuclear Measurements\label{tab:nuc}}
\tablewidth{0pt}
\tablehead{
\colhead{NAME} & \colhead{6.2~$\mu$m} & \colhead{7.7~$\mu$m\tablenotemark{a}} & \colhead{8.6~$\mu$m} & \colhead{11.3~$\mu$m\tablenotemark{b}} & \colhead{12.7~$\mu$m\tablenotemark{c}} & \colhead{\neii} & \colhead{H$_2$ S(3)} }
\startdata
 IC3639  &  1.15$\pm$0.05e-06  &  5.19$\pm$0.79e-06  &  3.41$\pm$0.54e-07  &  1.38$\pm$0.05e-06  &  4.94$\pm$0.54e-07  &  3.11$\pm$0.05e-07  &  4.45$\pm$0.79e-08 \\
NGC1058  &  1.47$\pm$0.10e-07  &  4.81$\pm$0.64e-07  &  9.52$\pm$0.69e-08  &  1.91$\pm$0.06e-07  &  9.12$\pm$0.88e-08  &  9.00$\pm$1.09e-09  &  5.22$\pm$1.28e-09 \\
NGC1097  &  9.24$\pm$0.31e-07  &  3.69$\pm$0.19e-06  &  6.90$\pm$0.22e-07  &  1.48$\pm$0.03e-06  &  8.07$\pm$0.21e-07  &  1.54$\pm$0.02e-07  &  8.36$\pm$0.68e-08 \\
NGC1241  &  5.38$\pm$0.83e-07  &  1.95$\pm$0.44e-06  &  3.50$\pm$0.44e-07  &  5.86$\pm$0.71e-07  &  2.67$\pm$0.35e-07  &  7.89$\pm$0.48e-08  &  3.65$\pm$1.61e-08 \\
NGC1365  &  4.77$\pm$0.07e-06  &  1.98$\pm$0.05e-05  &  3.10$\pm$0.05e-06  &  3.85$\pm$0.04e-06  &  3.00$\pm$0.06e-06  &  5.12$\pm$0.05e-07  &  1.11$\pm$0.07e-07 \\
NGC1433  &  3.58$\pm$0.25e-07  &  8.93$\pm$1.79e-07  &  2.50$\pm$0.18e-07  &  6.08$\pm$0.15e-07  &  2.66$\pm$0.23e-07  &  4.96$\pm$0.21e-08  &  3.34$\pm$0.62e-08 \\
NGC1566  &  8.12$\pm$0.25e-07  &  3.16$\pm$0.14e-06  &  5.14$\pm$0.20e-07  &  1.06$\pm$0.01e-06  &  4.87$\pm$0.18e-07  &  7.46$\pm$0.20e-08  &  5.01$\pm$0.32e-08 \\
NGC2273  &  2.54$\pm$0.03e-06  &  9.23$\pm$0.14e-06  &  1.45$\pm$0.02e-06  &  2.69$\pm$0.03e-06  &  1.08$\pm$0.02e-06  &  3.20$\pm$0.03e-07  &  1.04$\pm$0.04e-07 \\
NGC2639  &  9.29$\pm$0.97e-08  &  2.13$\pm$0.61e-07  &  4.19$\pm$0.66e-08  &  1.34$\pm$0.05e-07  &  8.02$\pm$0.71e-08  &  7.55$\pm$0.09e-08  &  1.47$\pm$0.13e-08 \\
NGC2992  &  1.55$\pm$0.05e-06  &  6.82$\pm$0.35e-06  &  8.24$\pm$0.36e-07  &  1.56$\pm$0.03e-06  &  7.80$\pm$0.43e-07  &  3.49$\pm$0.04e-07  &  5.99$\pm$0.52e-08 \\
NGC3079  &  1.82$\pm$0.03e-05  &  7.32$\pm$0.08e-05  &  1.19$\pm$0.03e-05  &  1.15$\pm$0.03e-05  &  6.68$\pm$0.07e-06  &  9.66$\pm$0.10e-07  &  5.04$\pm$0.26e-07 \\
NGC3185  &  1.11$\pm$0.05e-06  &  3.51$\pm$0.26e-06  &  6.99$\pm$0.40e-07  &  1.00$\pm$0.04e-06  &  4.57$\pm$0.34e-07  &  7.56$\pm$0.36e-08  &  4.19$\pm$0.98e-08 \\
NGC3227  &  2.91$\pm$0.03e-06  &  9.99$\pm$0.15e-06  &  1.24$\pm$0.01e-06  &  3.24$\pm$0.03e-06  &  1.35$\pm$0.01e-06  &  5.17$\pm$0.05e-07  &  2.03$\pm$0.03e-07 \\
NGC3735  &  5.92$\pm$0.42e-07  &  3.39$\pm$0.34e-06  &  5.61$\pm$0.28e-07  &  7.59$\pm$0.30e-07  &  4.47$\pm$0.38e-07  &  5.84$\pm$0.37e-08  &  2.44$\pm$0.69e-08 \\
NGC4051  &  8.84$\pm$0.54e-07  &  4.42$\pm$0.31e-06  &  3.96$\pm$0.32e-07  &  1.12$\pm$0.03e-06  &  5.67$\pm$0.52e-07  &  1.04$\pm$0.06e-07  &  8.09$\pm$0.77e-08 \\
NGC4258  &  3.40$\pm$0.07e-07  &  1.45$\pm$0.07e-06  &  1.36$\pm$0.04e-07  &  3.38$\pm$0.03e-07  &  1.60$\pm$0.06e-07  &  7.43$\pm$0.07e-08  &  6.58$\pm$0.14e-08 \\
NGC4501  &  1.09$\pm$0.05e-07  &  3.51$\pm$0.35e-07  &  8.02$\pm$0.37e-08  &  3.05$\pm$0.04e-07  &  1.50$\pm$0.03e-07  &  4.15$\pm$0.04e-08  &  6.37$\pm$0.25e-08 \\
NGC4639  &  9.15$\pm$0.56e-08  &  2.33$\pm$0.30e-07  &  3.47$\pm$0.35e-08  &  1.67$\pm$0.03e-07  &  6.65$\pm$0.52e-08  &  9.47$\pm$0.57e-09  &  9.22$\pm$0.69e-09 \\
NGC4945  &  7.99$\pm$0.08e-05  &  3.80$\pm$0.04e-04  &  3.91$\pm$0.04e-05  &  2.51$\pm$0.03e-05  &  2.81$\pm$0.03e-05  &  5.59$\pm$0.06e-06  &  3.24$\pm$0.71e-08 \\
NGC5005  &  1.40$\pm$0.06e-06  &  5.39$\pm$0.31e-06  &  1.20$\pm$0.04e-06  &  3.38$\pm$0.05e-06  &  1.26$\pm$0.03e-06  &  3.27$\pm$0.04e-07  &  4.13$\pm$0.22e-07 \\
NGC5033  &  6.04$\pm$0.40e-07  &  2.25$\pm$0.25e-06  &  4.07$\pm$0.21e-07  &  1.10$\pm$0.03e-06  &  5.77$\pm$0.19e-07  &  1.23$\pm$0.03e-07  &  4.00$\pm$0.57e-08 \\
NGC5135  &  5.65$\pm$0.06e-06  &  2.07$\pm$0.02e-05  &  3.62$\pm$0.04e-06  &  4.76$\pm$0.05e-06  &  2.58$\pm$0.03e-06  &  8.04$\pm$0.08e-07  &  1.22$\pm$0.04e-07 \\
NGC5194  &  2.57$\pm$0.32e-07  &  6.23$\pm$1.25e-07  &  1.12$\pm$0.19e-07  &  6.82$\pm$0.16e-07  &  2.85$\pm$0.17e-07  &  1.68$\pm$0.02e-07  &  1.32$\pm$0.09e-07 \\
NGC5395  &  6.69$\pm$0.43e-08  &  2.45$\pm$0.34e-07  &  4.14$\pm$0.38e-08  &  1.22$\pm$0.05e-07  &  5.88$\pm$0.45e-08  &  4.44$\pm$0.50e-09  &  6.12$\pm$1.16e-09 \\
NGC5427  &  2.23$\pm$0.06e-07  &  1.12$\pm$0.06e-06  &  1.49$\pm$0.05e-07  &  2.68$\pm$0.04e-07  &  1.43$\pm$0.08e-07  &  4.03$\pm$0.06e-08  &  1.32$\pm$0.11e-08 \\
NGC5643  &  1.10$\pm$0.09e-06  &  4.32$\pm$0.87e-06  &  4.27$\pm$1.17e-07  &  1.93$\pm$0.09e-06  &  9.62$\pm$1.02e-07  &  1.66$\pm$0.10e-07  &  7.28$\pm$2.05e-08 \\
NGC5728  &  9.91$\pm$0.23e-07  &  5.03$\pm$0.17e-06  &  7.19$\pm$0.17e-07  &  2.10$\pm$0.02e-06  &  8.19$\pm$0.17e-07  &  2.18$\pm$0.07e-07  &  1.55$\pm$0.07e-07 \\
NGC6221  &  9.07$\pm$0.09e-06  &  3.02$\pm$0.03e-05  &  4.92$\pm$0.05e-06  &  6.50$\pm$0.07e-06  &  3.88$\pm$0.05e-06  &  1.77$\pm$0.02e-06  &  1.62$\pm$0.09e-07 \\
NGC6951  &  1.87$\pm$0.05e-06  &  6.39$\pm$0.26e-06  &  1.30$\pm$0.04e-06  &  2.04$\pm$0.06e-06  &  9.54$\pm$0.29e-07  &  2.44$\pm$0.04e-07  &  8.03$\pm$1.14e-08 \\
NGC7130  &  2.71$\pm$0.05e-06  &  1.08$\pm$0.03e-05  &  1.71$\pm$0.05e-06  &  2.48$\pm$0.06e-06  &  1.37$\pm$0.06e-06  &  5.21$\pm$0.07e-07  &  6.38$\pm$1.11e-08 \\
NGC7314  &  1.71$\pm$0.23e-07  &  1.20$\pm$0.17e-06  &  1.09$\pm$0.17e-07  &  2.57$\pm$0.11e-07  &  1.33$\pm$0.14e-07  &  6.41$\pm$0.15e-08  &  2.50$\pm$0.36e-08 \\
NGC7469  &  8.23$\pm$0.08e-06  &  2.93$\pm$0.06e-05  &  4.28$\pm$0.04e-06  &  5.41$\pm$0.05e-06  &  3.29$\pm$0.07e-06  &  1.16$\pm$0.01e-06  &  1.32$\pm$0.05e-07 \\
NGC7496  &  2.18$\pm$0.05e-06  &  6.48$\pm$0.25e-06  &  1.15$\pm$0.03e-06  &  1.42$\pm$0.03e-06  &  7.30$\pm$0.32e-07  &  2.77$\pm$0.04e-07  &  6.14$\pm$0.93e-08 \\
NGC7582  &  2.26$\pm$0.05e-06  &  8.55$\pm$0.26e-06  &  2.58$\pm$0.04e-06  &  4.10$\pm$0.04e-06  &  1.55$\pm$0.03e-06  &  3.60$\pm$0.04e-07  &  1.37$\pm$0.10e-07 \\
NGC7590  &  3.01$\pm$0.57e-07  &  1.19$\pm$0.31e-06  &  2.72$\pm$0.36e-07  &  4.49$\pm$0.48e-07  &  1.95$\pm$0.29e-07  &  2.72$\pm$0.30e-08  &  2.18$\pm$1.10e-08 \\
\enddata
\tablecomments{Measurements are in units of W~m$^{-2}$~sr$^{-1}$.}
\tablenotetext{a}{Consists of sub-features at 7.42, 7.60, and 7.85~$\mu$m.}
\tablenotetext{b}{Consists of sub-features at 11.23 and 11.33~$\mu$m.}
\tablenotetext{c}{Consists of sub-features at 12.62 and 12.69~$\mu$m.}
\end{deluxetable}
  
\begin{deluxetable}{lllccccc}
\tabletypesize{\scriptsize}
\tablecaption{Off-Nuclear Measurements\label{tab:off}}
\tablewidth{0pt}
\tablehead{
\colhead{NAME} & \colhead{RA} & \colhead{Dec} & \colhead{6.2~$\mu$m} & \colhead{7.7~$\mu$m\tablenotemark{a}} & \colhead{8.6~$\mu$m} & \colhead{11.3~$\mu$m\tablenotemark{b}} & \colhead{12.7~$\mu$m\tablenotemark{c}} }
\startdata
 IC3639  &  12:40:53.13  &  $-$36:45:10.3  &  4.24$\pm$0.32e-07  &  1.63$\pm$0.22e-06  &  1.94$\pm$0.28e-07  &  2.84$\pm$0.20e-07  &  1.63$\pm$0.32e-07 \\
NGC1097  &  02:46:19.06  &  $-$30:16:20.0  &  4.26$\pm$0.07e-06  &  1.39$\pm$0.04e-05  &  2.63$\pm$0.05e-06  &  2.86$\pm$0.05e-06  &  1.62$\pm$0.03e-06 \\
NGC1365  &  03:33:36.71  &  $-$36:08:18.0  &  8.53$\pm$0.09e-06  &  3.37$\pm$0.04e-05  &  6.68$\pm$0.08e-06  &  6.24$\pm$0.07e-06  &  4.15$\pm$0.05e-06 \\
NGC1566  &  04:20:02.13  &  $-$54:56:37.1  &  2.69$\pm$0.18e-07  &  8.33$\pm$1.14e-07  &  1.16$\pm$0.13e-07  &  1.33$\pm$0.10e-07  &  7.92$\pm$1.39e-08 \\
NGC2992  &  09:45:42.07  &  $-$14:19:29.4  &  1.05$\pm$0.10e-06  &  3.59$\pm$0.75e-06  &  6.48$\pm$0.84e-07  &  9.31$\pm$1.03e-07  &  6.07$\pm$0.47e-07 \\
NGC3079  &  10:01:57.49  &  +55:40:58.5  &  2.61$\pm$0.10e-06  &  9.02$\pm$0.30e-06  &  1.58$\pm$0.09e-06  &  1.80$\pm$0.08e-06  &  9.90$\pm$0.27e-07 \\
NGC3227  &  10:23:30.87  &  +19:51:43.1  &  1.04$\pm$0.08e-07  &  4.36$\pm$0.43e-07  &  7.23$\pm$0.46e-08  &  1.51$\pm$0.07e-07  &  7.30$\pm$0.49e-08 \\
NGC4258  &  12:18:59.31  &  +47:18:24.8  &  4.44$\pm$0.05e-07  &  1.44$\pm$0.02e-06  &  2.60$\pm$0.03e-07  &  2.87$\pm$0.03e-07  &  1.51$\pm$0.04e-07 \\
NGC4501  &  12:32:00.42  &  +14:25:25.2  &  2.97$\pm$0.08e-07  &  1.09$\pm$0.03e-06  &  1.92$\pm$0.05e-07  &  2.43$\pm$0.06e-07  &  1.33$\pm$0.04e-07 \\
NGC4945  &  13:05:28.26  &  $-$49:27:39.6  &  2.24$\pm$0.06e-06  &  7.69$\pm$0.11e-06  &  1.61$\pm$0.07e-06  &  1.72$\pm$0.06e-06  &  9.13$\pm$0.19e-07 \\
NGC5005  &  13:10:56.89  &  +37:03:24.8  &  4.87$\pm$0.65e-07  &  1.98$\pm$0.50e-06  &  3.28$\pm$0.32e-07  &  4.59$\pm$0.25e-07  &  2.71$\pm$0.34e-07 \\
NGC5033  &  13:13:27.87  &  +36:35:25.6  &  9.14$\pm$0.28e-07  &  3.47$\pm$0.15e-06  &  6.73$\pm$0.34e-07  &  7.55$\pm$0.23e-07  &  4.72$\pm$0.21e-07 \\
NGC5135  &  13:25:44.60  &  $-$29:50:08.6  &  1.90$\pm$0.13e-07  &  8.02$\pm$0.97e-07  &  1.31$\pm$0.09e-07  &  2.24$\pm$0.07e-07  &  1.27$\pm$0.25e-07 \\
NGC5194  &  13:29:50.36  &  +47:11:36.0  &  7.64$\pm$0.25e-07  &  2.82$\pm$0.11e-06  &  4.08$\pm$0.24e-07  &  5.82$\pm$0.19e-07  &  3.60$\pm$0.20e-07 \\
NGC5395  &  13:58:38.82  &  +37:25:38.2  &  2.49$\pm$0.04e-07  &  8.21$\pm$0.31e-07  &  1.47$\pm$0.04e-07  &  1.61$\pm$0.03e-07  &  8.41$\pm$0.56e-08 \\
NGC5427  &  14:03:26.11  &  $-$06:01:43.2  &  2.64$\pm$0.07e-07  &  1.02$\pm$0.05e-06  &  1.87$\pm$0.05e-07  &  2.24$\pm$0.04e-07  &  1.24$\pm$0.07e-07 \\
NGC6221  &  16:52:46.03  &  $-$59:13:08.8  &  1.05$\pm$0.04e-06  &  3.04$\pm$0.19e-06  &  6.22$\pm$0.25e-07  &  1.21$\pm$0.04e-06  &  6.46$\pm$0.31e-07 \\
NGC7130  &  21:48:19.38  &  $-$34:56:56.1  &  1.31$\pm$0.03e-06  &  4.35$\pm$0.13e-06  &  8.07$\pm$0.25e-07  &  1.03$\pm$0.03e-06  &  5.14$\pm$0.14e-07 \\
NGC7314  &  22:35:46.89  &  $-$26:03:13.7  &  1.11$\pm$0.13e-07  &  4.55$\pm$1.01e-07  &  8.01$\pm$1.30e-08  &  1.00$\pm$0.10e-07  &  5.26$\pm$1.40e-08 \\
NGC7582  &  23:18:22.64  &  $-$42:21:57.7  &  4.14$\pm$0.29e-07  &  1.82$\pm$0.19e-06  &  2.29$\pm$0.24e-07  &  3.36$\pm$0.18e-07  &  1.76$\pm$0.31e-07 \\
NGC7590  &  23:18:55.05  &  $-$42:14:28.0  &  5.19$\pm$0.53e-07  &  1.78$\pm$0.37e-06  &  2.81$\pm$0.29e-07  &  3.58$\pm$0.22e-07  &  1.79$\pm$0.29e-07 \\
\enddata
\tablecomments{Measurements are in units of W~m$^{-2}$~sr$^{-1}$.}
\tablenotetext{a}{Consists of sub-features at 7.42, 7.60, and 7.85~$\mu$m.}
\tablenotetext{b}{Consists of sub-features at 11.23 and 11.33~$\mu$m.}
\tablenotetext{c}{Consists of sub-features at 12.62 and 12.69~$\mu$m.}
\end{deluxetable}

\begin{deluxetable}{llll}
\tablecaption{Statistical Tests\label{tab:stats}}
\tablewidth{0pt}

\tablehead{ \colhead{ratio} & \colhead{Seyferts v. SINGS} &
\colhead{Seyferts v. off nuclear} & \colhead{SINGS v. off nuclear}}

\startdata
6/11  & $\bf{5\times10^{-4}}$  & $\bf{5\times10^{-5}}$  & 0.682 \\
7/11  & {\bf 0.003}            & {\bf 0.001}            & 0.074 \\
8/11  & $\bf{9\times10^{-4}}$  & $\bf{2\times10^{-4}}$  & 0.063 \\
6/7   & 0.230                  & 0.447                  & 0.888 \\
6/8   & 0.489                  & 0.347                  & 0.689 \\
7/8   & 0.108                  & 0.303                  & 0.374 \\
\enddata
\tablecomments{Values correspond to probabilities from two-sample K-S tests.}
\end{deluxetable}

\end{document}